\documentclass[aps,prb,twocolumn,amsmath,amssymb,eqsecnum]{revtex4-1}

\usepackage{amsmath}
\usepackage{amssymb}
\usepackage[pdftex]{color}
\usepackage{graphicx}
\usepackage{dcolumn} 
\usepackage{bm} 
\usepackage{longtable}
\usepackage{ulem}

\usepackage{enumitem}

\usepackage{color}
\usepackage{graphicx}
\usepackage{curves}
\usepackage{epic}
\usepackage{wasysym}
\usepackage{epsf, times}
\usepackage{subfigure}
\usepackage{eufrak}
\usepackage{amsthm}

\newcommand{\bs}[1]{{\boldsymbol{#1}}}

\newcommand{\ket}[1]{{|\,{#1}\,\rangle}}
\newcommand{\bra}[1]{{\langle\,{#1}\,|}}

\newcommand{\s}{{\sigma}}

\begin{document}

\author{Luiz H. Santos}
\affiliation{Perimeter Institute for Theoretical Physics, Waterloo, ON, N2L 2Y5, Canada}

\title{
     Rokhsar-Kivelson Models of Bosonic Symmetry-Protected Topological States
      }

\begin{abstract}

A platform for constructing microscopic Hamiltonians describing bosonic symmetry-protected topological (SPT) states is presented. 
The Hamiltonians we consider are examples of frustration-free Rokhsar-Kivelson models, which
are known to be in one-to-one correspondence with classical stochastic systems in the
same spatial dimensionality.
By exploring this classical-quantum mapping, we are able to construct a large class of microscopic models which, in a closed manifold, have a non-degenerate gapped symmetric ground state describing the universal properties of SPT states.
Examples of one and two dimensional SPT states which illustrate our approach are discussed.

\end{abstract}


\maketitle



\section{Introduction}
\label{sec:Introduction}

The $SO(3)$ symmetric spin-1 anti-ferromagnetic Heisenberg chain, which was shown by Haldane~\cite{Haldane-1983-a,Haldane-1983-b,Haldane-1985} to have a symmetry preserving gapped ground state, provides the oldest known example of a bosonic symmetry-protected topological (SPT) state in one dimension.
The topological character of this state is captured by a topological $\theta$-term present in the non-linear sigma model effective action describing long-wavelength degrees of freedom.~\cite{Haldane-1983-a,Haldane-1983-b,Haldane-1985}

An insightful account of the properties of this anti-ferromagnetic spin chain was given by Affleck, Kennedy, Lieb, and Tasaki (AKLT), who constructed a Hamiltonian in the same phase as the Heisenberg model, where the $S=1$ spins emerge from the composition of underlying $S=1/2$ degrees of freedom.~\cite{Affleck-1987,Affleck-1988}
With periodic boundary conditions, the AKLT model has a non-degenerate ground state that does not break any symmetries and is separated from the first excited state by a finite gap.
With open boundary conditions, on the other hand, the AKLT model makes it manifest that each edge supports a ``free" $S=1/2$ degree of freedom contributing to a 2-fold degeneracy per edge. 
Interestingly, while the Hamiltonian is $SO(3)$ symmetric and the bulk degrees of freedom transform linearly under this symmetry, the effective $S=1/2$ spins on the edges transform projectively under the action of the spin rotation symmetry: an initial $S=1/2$ spin state rotated by $360$ degrees about an arbitrary axis is mapped into itself up to a minus sign. 
The inter-connection among the topological $\theta$-term action, the ground state degeneracy with open boundary conditions and the projective representation of the global symmetry on the edge degrees of freedom makes this system a non-trivial gapped phase of matter.

Inspired by the example of the anti-ferromagnetic chain, there have been recent proposals to classify gapped 
SPT phases of matter protected by a global symmetry $G$ using various mathematical frameworks such as group cohomology,~\cite{Chen-2012-a,Chen-2013} which generalizes the concept of projective representations,
topological field theories~\cite{Lu-2012,Chen-2012-b,Vishwanath-2013,Metliski-2013-a,Sule-2013,C-Wang-2014,Ye-2013,Santos-2014,J-Wang-2014}
and non-linear sigma models in the presence of a topological $\theta$-term action compatible with the 
global symmetry $G$.~\cite{Bi-2013,Xu-2013}
Recently, a number of microscopic models of bosonic SPT states have been 
studied, which help to shed light on the role played by physical interactions in bringing about SPT phases.~\cite{Chen-2012-a,Chen-2013,Chen-2012-b,Santos-2014,J-Wang-2014,Chen-2011,Levin-2012,Chen-2014,Burnell-2013,Fidkowski-2013,Lu-2014,Chen-2014-2,Geraedts-2014}

The purpose of this paper is to provide a framework for constructing microscopic models capable of describing bosonic SPT states. 
As we shall see, some of the exactly solvable models 
previously studied~\cite{Levin-2012,Chen-2014,Geraedts-2014}
will be identified as special cases of a large class of models to be constructed here.
We shall also be able to construct parent Hamiltonians for two dimensional
$
\mathbb{Z}_{2}\times\mathbb{Z}_{2}
$
paramagnets,
whose effective edge theory was shown in Ref.~\onlinecite{J-Wang-2014}
to be in direct relation to non-trivial $3$-cocycles.

The classes of gapped bosonic insulators protected a by global symmetry $G$ that we shall be concerned with have, on a 
$\mathrm{d}$-dimensional closed manifold, a non-degenerate ground state 
\begin{equation}
\label{eq: SPT ground state}
\begin{split}
\ket{\Psi_{G}}
=
\frac{1}{\sqrt{Z(\beta)}}\,
\sum_{ s }\,
e^
{
-\frac{\beta}{2}\,E_{G}(s)
+
i\,W_{G}(s)
}
\,
\ket{s}
\,,
\end{split}•
\end{equation}•
where $\{ \ket{s} \}$ denotes an orthonormal many-body basis, $E_{G}(s) \in \mathbb{R}$ is a non-universal local function related to the decay of correlations of local operators in the ground state and the phase $W_{G}(s) \in \mathbb{R}$ is a universal piece that endows the ground state~(\ref{eq: SPT ground state}) with its non-trivial topological properties. 
$W_{G}(s)$ plays, at the microscopic level considered here, a role analogous to the topological $\theta$-term.~\cite{Chen-2012-a,Chen-2013,Bi-2013,Xu-2013}

When $W_{G}(s)= 0$, one obtains from Eq.~(\ref{eq: SPT ground state}) 
the nodeless ground state 
\begin{equation}
\label{eq: RK ground state}
\begin{split}
\ket{\Phi_{G}}
=
\frac{1}{\sqrt{Z(\beta)}}\,
\sum_{ s }\,
e^
{
-\frac{\beta}{2}\,E_{G}(s)
}
\,
\ket{s}
\,.
\end{split}•
\end{equation}•
The form of the ground state Eq.~(\ref{eq: RK ground state}) is very appealing; for 
equal time correlation functions of operators in the diagonal representation $\{ \ket{s} \}$,
\begin{equation}
\label{eq: Correlation function of diagonal operators}
\begin{split}
&\,
\bra{\Phi_{G}}\,
\hat{\mathcal{O}}_{a}(\hat{s})\,
\hat{\mathcal{O}}_{b}(\hat{s})\,
\ket{\Phi_{G}}
=
\sum_{  s  }\,
\mathcal{O}_{a}(s)\,
\mathcal{O}_{b}(s)\,
\frac
{
e^
{
-\beta\,E_{G}(s)
}
}
{
Z(\beta)
}
\,,
\end{split}•
\end{equation}•
can be interpreted as equal time correlation functions
of an equilibrium 
$\mathrm{d}$-dimensional
statistical mechanical system with classical configurations $\{ s \}$,
each one occurring with probability
$
p^{(0)}_{s}
=
e^
{
-\beta\,E_{G}(s)
}
/
Z(\beta)
$,
where the real parameter $\beta$ acquires the natural interpretation of an effective inverse temperature and the normalization factor of the ground state,
\begin{equation}
\label{eq: Normalization of the ground state}
\begin{split}
Z(\beta)
=
\sum_{ s }\,
e^
{
-\beta\,E_{G}(s)
}
\,,
\end{split}•
\end{equation}•
is interpreted as the partition function of the classical system in ``thermal" equilibrium.
Hence, if the associated classical model described by the partition function 
Eq.~(\ref{eq: Normalization of the ground state}) is in the ``disordered" phase, 
then typical correlation functions of local operators distant by 
$|\bs{r}|$ behave as
$
e^
{
-|\bs{r}|/\xi
}
$,
for some finite correlation length
$
\xi
$,
and the representation~(\ref{eq: RK ground state}) can be associated with a 
quantum many-body ground state in its gapped phase.

In fact the foregoing classical-quantum correspondence can be made more precise.~\cite{Rokhsar-1988,Henley-1997,Ardonne-2004,Henley-2004,Castelnovo-2005}
Eq.~(\ref{eq: RK ground state}) is recognized as the zero energy ground state of a class of 
quantum dimer-like models at the so-called Rokhsar-Kivelson (RK) point.~\cite{Rokhsar-1988}
In Ref.~\onlinecite{Rokhsar-1988}, it was noted that dimer-dimer correlation functions of the square lattice quantum dimer model at the RK point can be computed exactly from the 
corresponding classical dimer problem.~\cite{Fisher-1963}
In Ref.~\onlinecite{Ardonne-2004}, Ardonne, Fendley and Fradkin have established 
the quantum-classical correspondence to more general classes of RK Hamiltonians beyond dimer models.
In Ref.~\onlinecite{Henley-2004},
Henley has observed that any stochastic classical system described by a real transition rate matrix $M$
can be interpreted, via a similarity transformation, as an RK Hamiltonian.
In Ref.~\onlinecite{Castelnovo-2005}, Castelnovo, Chamon, Mudry and Pujol have shown that the reverse is true, namely, that given a quantum RK Hamiltonian in a ``preferred basis" [the basis in which the ground state is expressed as a linear combination of 
same-sign coefficients as in Eq.~(\ref{eq: RK ground state})], there exists an associated stochastic classical model whose spectrum of relaxation rates is (up to an overall minus sign) the same as the energy spectrum of the quantum RK Hamiltonian and whose equilibrium probability distribution is the square of the coefficients in the expansion of the RK quantum ground state 
Eq.~(\ref{eq: RK ground state}).

In light of the above arguments, if one considers the configurations $\{ s \} $ to be made of spins on a lattice, then the ground state
Eq.~(\ref{eq: RK ground state}) offers a natural representation of a paramagnetic
state, provided the corresponding classical system is chosen to have a spectrum of relaxation rates with a finite gap and correlation functions of local 
operators, Eq.~(\ref{eq: Correlation function of diagonal operators}), exhibiting
short-range behavior.

As for the role played by symmetries, we now let the quantum system be invariant under a global symmetry group $G$, whose action on the basis
$
\ket{ s }
$ 
is represented by
\begin{equation}
\label{eq: Action of the symmetry on the basis elements}
\widehat{S}_{G}\,\ket{ s }
=
\ket{ g\,s }
\,.
\end{equation}•
It is then clear that the ground state Eq.~(\ref{eq: RK ground state}) is a unique and
$G$ invariant state provided the local ``classical energy" 
$
E_{G}(s)
$
is symmetric under the transformation
$
\{ s \} 
\rightarrow
\{ g\,s \} 
$:
\begin{equation}
\label{eq: Transformation of E under G}
\begin{split}
E_{G}(g\,s)
=
E_{G}(s)
\,.
\end{split}•
\end{equation}•

The central point of this paper is the observation, which will be supported by concrete examples, that in a \textit{closed} manifold, the SPT 
ground state Eq.~(\ref{eq: SPT ground state}) can be obtained from the 
trivial insulator ground state Eq.~(\ref{eq: RK ground state}) via a \textit{global} symmetry-preserving unitary transformation
$
\mathbb{W}_{G}
$,
whose action on the many-body basis $\{ \ket{s} \}$ is 
\begin{subequations}
\label{eq: Definition of unitary mapping}
\begin{equation}
\label{eq: Action of W on the basis elements}
\begin{split}
\mathbb{W}_{G}\,\ket{s}
:=
e^
{
i\,W_{G}(s)
}
\,
\ket{s}
\,,
\end{split}•
\end{equation}•
hence,
\begin{equation}
\label{eq: Unitary mapping between Psi and Phi}
\begin{split}
\ket{\Psi_{G}}
=
\mathbb{W}_{G}\,
\ket{\Phi_{G}}
\,.
\end{split}•
\end{equation}•
\end{subequations}•

For the SPT ground state Eq.~(\ref{eq: SPT ground state}) to be invariant under the symmetry $G$, it is required that
\begin{subequations}
\begin{equation}
\label{eq: Transformation of E and W under G unitary}
\begin{split}
&\,
E_{G}(g\,s)
=
E_{G}(s)
\,,
\\
&\,
W_{G}(g\,s)
=
W_{G}(s)
~
\textrm{mod}~2\pi
\,,
\end{split}•
\end{equation}•
if $G$ is an unitary global symmetry, and
\begin{equation}
\label{eq: Transformation of E and W under G anti-unitary}
\begin{split}
&\,
E_{G}(g\,s)
=
E_{G}(s)
\,,
\\
&\,
W_{G}(g\,s)
=
- W_{G}(s)
~
\textrm{mod}~2\pi
\,,
\end{split}•
\end{equation}•
\end{subequations}•
if $G$ is an anti-unitary global symmetry.

Now let $H_{G}$ and $\mathcal{H}_{G}$ be,
respectively, the quantum Hamiltonians whose non-degenerate ground states are
$\ket{\Phi_{G}}$ and $\ket{\Psi_{G}}$.
Then the unitary mapping Eq.~(\ref{eq: Unitary mapping between Psi and Phi}) establishes
\begin{equation}
\label{eq: Unitary mapping between Hamiltonians}
\mathcal{H}_{G}
=
\mathbb{W}_{G}
\,
H_{G}
\,
\mathbb{W}_{G}^{-1}
\,.
\end{equation}•
Starting, thus, from a parent Hamiltonian $H_{G}$ for the trivial gapped state
Eq.~(\ref{eq: RK ground state}), 
one can construct a parent Hamiltonian $\mathcal{H}_{G}$ for the SPT ground state 
Eq.~(\ref{eq: SPT ground state})
using Eq.~(\ref{eq: Unitary mapping between Hamiltonians}), if the unitary transformation connecting the two ground states, 
Eq.~(\ref{eq: Definition of unitary mapping}), is known.

This paper is organized as follows.
In Sec.~(\ref{sec: Review of the classical-quantum mapping}) we review
the relevant points about the mapping between stochastic classical systems 
and quantum RK Hamiltonians. 
In Sec.~(\ref{sec: SPT states in one dimension}) we construct parent Hamiltonians of one dimensional bosonic SPT states with $\mathbb{Z}_{n} \times \mathbb{Z}_{n}$ symmetry, where we shall use the concept of entanglement spectrum degeneracy to determiner the unitary transformation
Eq.~(\ref{eq: Definition of unitary mapping})
that maps the trivial ground state to other $n-1$ topological phases.
In Sec.~(\ref{sec: Z2T SPT state in one dimension}) we discuss the one 
dimensional SPT state with anti-unitary time-reversal symmetry 
$\mathbb{Z}^{T}_{2}$.
In Sec.~(\ref{SPT states in two dimensions}) we construct two dimensional microscopic models that account for all the $8$ possible classes of $\mathbb{Z}_{2} \times \mathbb{Z}_{2}$ SPT paramagnets.
We show that the $\mathbb{Z}_{2} \times \mathbb{Z}_{2}$ symmetry transformation projected onto the one dimensional edge acquires a non-onsite form, which was studied in Ref.~\onlinecite{J-Wang-2014} in connection with non-trivial $3$-cocycles in the group cohomology.
Finally we draw some conclusions and point to future directions in 
Sec.~(\ref{sec:Conclusions and future directions}).

\section{Classical-quantum mapping}
\label{sec: Review of the classical-quantum mapping}

In order to make the discussion self-contained, we review the essential aspects of the relationship between quantum RK Hamiltonians and stochastic classical  systems.~\cite{Henley-2004,Castelnovo-2005} The content of 
Secs.~(\ref{sec: Rokhsar-Kivelson Hamiltonians})
and~(\ref{sec: Relation between quantum RK Hamiltonians and stochastic classical systems})
closely follows Ref.~\onlinecite{Castelnovo-2005}, which the reader may consult for further details. 
In Sec.~(\ref{sec: SPT-RK Hamiltonians}) we give the general form of the SPT-RK Hamiltonians describing bosonic
SPT states in $\mathrm{d}$-dimensional space obtained from the unitary mapping
Eq.~(\ref{eq: Unitary mapping between Hamiltonians}).

\subsection{Rokhsar-Kivelson Hamiltonians}
\label{sec: Rokhsar-Kivelson Hamiltonians}

A quantum RK Hamiltonian satisfies three properties~\cite{Castelnovo-2005}:

\begin{enumerate}
\item
The orthonormal elements of the basis
$
\Sigma
=
\{
\ket
{
s
}
\}
$,
which span the Hilbert space,
\begin{equation}
\label{eq: Definition of orthonormal basis sigma}
\begin{split}
\bra{s} s' \rangle
=
\delta_{s,s'}
\,,
\quad
\openone
=
\sum_{s}\,
\ket{s}
\bra{s}
\,,
\end{split}•
\end{equation}•
form a countable set.

\item
The quantum Hamiltonian can be decomposed into a sum of positive-semidefinite projector-like Hermitian operators $P_{s,s'}$ 
[Eq.~(\ref{eq: Definition of the RK Hamiltonian})].
\item
The ground state [Eq.~(\ref{eq: RK ground state 2})] is annihilated by every $P_{s,s'}$ and the normalization constant of the ground state can be interpreted as the partition function of a classical system [Eq.~(\ref{eq: Definition of Z})].

\end{enumerate}•

The RK Hamiltonian takes the form
\begin{subequations}
\label{eq: Definition of the RK Hamiltonian}
\begin{equation}
\label{eq: RK Hamiltonian}
H_{\textrm{RK}}
=
\frac{1}{2}\,
\sum^{s \neq s\rq{}}_{(s,s')}\,
\omega_{s,s'}\,P_{s,s'}
\,,
\end{equation}•
where
\begin{equation}
\omega_{s,s'} \in \mathbb{R}
\,,
\quad
\omega_{s,s'} > 0
\,,
\end{equation}•
\begin{equation}
\label{eq: Definition of P}
\begin{split}
&\,P_{s,s'} 
=
-
\ket{s} \bra{s'}
-
\ket{s'} \bra{s}
\\
&\,
+
e^
{
-\frac{\beta}{2}\,
\left[
E(s') - E(s)
\right]
}
\,
\ket{s} \bra{s}
+
e^
{
-\frac{\beta}{2}\,
\left[
E(s) - E(s')
\right]
}
\,
\ket{s'} \bra{s'}
\,,
\end{split}•
\end{equation}•
\end{subequations}•
where $\beta \in \mathbb{R}$, $E(s) \in \mathbb{R}$ and
$P_{s,s'}$ is a projector-like positive semi-definite Hermitian operator satisfying
\begin{equation}
\begin{split}
P^{2}_{s,s'}
=
2\,
\cosh
{
\left\{
\frac{\beta}{2}\,
\left[
E(s')
-
E(s)
\right]
\right\}
}
\,
P_{s,s'}
\,.
\end{split}•
\end{equation}•

One easily verifies that the nodeless state
\begin{equation}
\label{eq: RK ground state 2}
\begin{split}
\ket{\Phi_\textrm{RK}}
=
\frac
{
1
}
{
\sqrt
{
Z(\beta)
}
}
\,
\sum_{s}\,
e^
{
-\frac{\beta}{2}\,E(s)
}
\,
\ket
{
s
}
\,,
\end{split}•
\end{equation}•
with normalization constant
\begin{equation}
\label{eq: Definition of Z}
Z(\beta)
=
\sum_{s}\,
e^
{
-\beta\,E(s)
}
\,,
\end{equation}•
satisfies
\begin{equation}
\begin{split}
P_{s,s'}\,\ket{\Phi_\textrm{RK}}
=
0\,,
\quad
\forall~(s,s')
\,.
\end{split}•
\end{equation}•
Thus Eq.~(\ref{eq: RK ground state 2}) is the ground state of the RK Hamiltonian Eq.~(\ref{eq: Definition of the RK Hamiltonian}) with energy zero.
The normalization constant Eq.~(\ref{eq: Definition of Z}) can be interpreted as the partition function of a classical system with classical energy $E(s)$ at the effective inverse temperature $\beta$.
In the ``infinite temperature" limit $\beta = 0$, the Boltzmann factors in the partition function tend to unity and $Z(0)$ counts the number of allowed configurations.

\subsection{Relation between quantum RK Hamiltonians and stochastic classical systems}
\label{sec: Relation between quantum RK Hamiltonians and stochastic classical systems}

In order to show the relation between quantum RK Hamiltonian and 
stochastic classical systems, one considers the real-valued matrix $M$ defined by~\cite{Henley-2004,Castelnovo-2005}
\begin{equation}
\label{eq: Definition of M}
M_{s s'}
\equiv
-
e^
{
-\frac{\beta}{2}\,
\left[
E(s)
-
E(s')
\right]
}
\,
\left(
H_\textrm{RK}
\right)_{ss'}
\,,
\end{equation}•
where
$
\left(
H_\textrm{RK}
\right)_{ss'}
=
\langle\,s\,|\,H_\textrm{RK}\,|\,s'\,\rangle
$
denotes the matrix elements of $H_\textrm{RK}$.
From the fact that $H_\textrm{RK}$ is a real and Hermitian operator, it follows that
\begin{equation}
\label{eq:	Symmetry of the RK Hamiltonian}
\left(
H_\textrm{RK}
\right)_{ss'}
=
\left(
H_\textrm{RK}
\right)_{s's}
\,
\end{equation}•
and, from Eqs.~(\ref{eq: Definition of the RK Hamiltonian}) and~(\ref{eq: Definition of M}), that the matrix $M$ satisfies:
\begin{subequations}
\label{eq: Properties of M}
\begin{equation}
\label{eq: Property of M (a)}
M_{s s'} > 0
\,,
\quad
\textrm{if}~
s \neq s'
\,,
\end{equation}•
and
\begin{equation}
\label{eq: Property of M (b)}
M_{s s}
=
-
\sum^{s' \neq s}_{s'}\,
M_{s' s}
\,.
\end{equation}•
\end{subequations}•

One then considers a classical system with a phase space formed
by configurations $\{ s \}$, which, as a function of time $\tau$, can be visited stochastically with probability $p_{s}(\tau)$ evolving according to the master equation 
\begin{equation}
\label{eq: Master equation}
\begin{split}
\frac{d\,p_{s}(\tau)}{d\,\tau}\,
&\,=
\sum_{s'}\,
M_{s s'}\,p_{s'}(\tau)
\\
&\,
=
\sum^{s' \neq s}_{s'}\,
\Big[
M_{s s'}\,p_{s'}(\tau)
-
M_{s' s}\,p_{s}(\tau)
\Big]
\,,
\end{split}•
\end{equation}•
where Eq.~(\ref{eq: Property of M (b)}) has been used to achieve the last equality of Eq.~(\ref{eq: Master equation}).
The first term on the r.h.s. accounts for the transitions out of the configurations $s'$ into the configuration $s$, while the second term on the r.h.s. accounts for transitions out of the configuration $s$ into the configurations $s'$.

Moreover Eqs.~(\ref{eq: Definition of M}) and~(\ref{eq: Symmetry of the RK Hamiltonian}) are easily seen to imply, for every pair of indices $(s,s')$,
\begin{subequations}
\label{eq: Detailed balance}
\begin{equation}
M_{s s'}\,p^{(0)}_{s'}
=
M_{s' s}\,p^{(0)}_{s}
\,,
\end{equation}•
where
\begin{equation}
p^{(0)}_{s}
\equiv
\frac
{
1
}
{
Z(\beta)
}
\,
e^
{
-\beta\,E(s)
}
\,.
\end{equation}•
\end{subequations}•
Eq.~(\ref{eq: Detailed balance}) implies the condition of detailed balance
on the matrix $M$ as well as that $p^{(0)}_{s}$ is the equilibrium probability distribution associated with the classical dynamics Eq.~(\ref{eq: Master equation}).

Denoting by $\lambda_{n}$ and $\psi^{(R;n)}_{s}$, respectively, the right-eigenvalues and right-eigenvectors of $M$,
\begin{equation}
\label{eq: Righ-eigenvalue equation of W}
\sum_{s'}\,
M_{s s'}\,
\psi^{(R;n)}_{s'}
=
\lambda_{n}\,
\psi^{(R;n)}_{s}
\,,
\end{equation}•
then the time dependent solution of Eq.~(\ref{eq: Master equation}) can be expressed as
\begin{equation}
\label{eq: Probability amplitudes as a function of time}
p_{s}(\tau)
=
\sum_{n}\,
a_{n}(0)\,
e^
{
\lambda_{n}\,\tau
}
\,
\psi^{(R;n)}_{s}
\,,
\end{equation}•
where $a_{n}(0)$ are coefficients determined by the initial conditions.

Since Eq.~(\ref{eq: Definition of M}) establishes, up to an overall minus sign,
a similarity transformation between $H_{\textrm{RK}}$ and $M$, the spectrum of
relaxation rates $\{ \lambda_{n} \}$ of $M$ and the energy spectrum 
$\{ \varepsilon_{n} \}$ of $H_{\textrm{RK}}$ are simply related:
\begin{equation}
\label{eq: Relation between eigenvalues of H and M}
\varepsilon_{n}
=
-\lambda_{n}
> 0
\,.
\end{equation}•

When the classical system whose dynamics is described by Eq.~(\ref{eq: Master equation}) has a spectrum of relaxation rates such that the largest characteristic time scale associated with the decay into the equilibrium configuration is finite in the thermodynamic limit, then it follows from Eq.~(\ref{eq: Relation between eigenvalues of H and M}) that the many-body energy spectrum of the quantum Hamiltonian possess a finite energy gap for excitations above ground state.

\subsection{SPT-RK Hamiltonians}
\label{sec: SPT-RK Hamiltonians}

We now give the general form of the SPT-RK Hamiltonians with ground state given by
Eq.~(\ref{eq: SPT ground state}).

Let the $\mathrm{d}$-dimensional quantum system, 
protected by global symmetry $G$, in its trivial phase be described by an
RK Hamiltonian 
$
H_{G}
$ 
of the form 
Eq.~(\ref{eq: Definition of the RK Hamiltonian}) 
with the non-degenerate ground state 
$
\ket
{
\Phi_{G}
}
$,
Eq.~(\ref{eq: RK ground state}),
where we impose the symmetry constraint
Eq.~(\ref{eq: Transformation of E under G})
upon $E_{G}(s)$.
Then the unitary transformation 
Eq.~(\ref{eq: Definition of unitary mapping})
yields the SPT ground state 
$
\ket
{
\Psi_{G}
}
$,
Eq.~(\ref{eq: SPT ground state}),
and 
Eq.~(\ref{eq: Unitary mapping between Hamiltonians})
yields the SPT-RK Hamiltonian
$
\mathcal{H}_{G}
$:
\begin{widetext}
\begin{subequations}
\label{eq: Definition of the SPT Hamiltonian}
\begin{equation}
\mathcal{H}_{G}
=
\mathbb{W}_{G}\,
H_{G}\,
\mathbb{W}_{G}^{-1}\,
=
\frac{1}{2}\,
\sum^{s' \neq s}_{(s,s')}\,
\omega_{s,s'}\,\mathcal{P}_{s,s'}
\,,
\end{equation}•
where
\begin{equation}
\omega_{s,s'} \in \mathbb{R}
\,,
\quad
\omega_{s,s'} > 0
\,,
\end{equation}•
\begin{equation}
\label{eq: Definition of mathcal P}
\begin{split}
&\,\mathcal{P}_{s,s'} 
=
-
e^
{
i\,
\left[
W_{G}(s)
-
W_{G}(s')
\right]
}
\,
\ket{s} \bra{s'}
-
e^
{
i\,
\left[
W_{G}(s')
-
W_{G}(s)
\right]
}
\,
\ket{s'} \bra{s}
+
e^
{
-\frac{\beta}{2}\,
\left[
E_{G}(s') - E_{G}(s)
\right]
}
\,
\ket{s} \bra{s}
+
e^
{
-\frac{\beta}{2}\,
\left[
E_{G}(s) - E_{G}(s')
\right]
}
\,
\ket{s'} \bra{s'}
\,.
\end{split}•
\end{equation}•
\end{subequations}•
\end{widetext}

Before we proceed to discuss specific SPT systems, we close this section with a few important remarks:

\begin{enumerate}[label=(\alph*)]
\item
In Secs.~(\ref{sec: Rokhsar-Kivelson Hamiltonians})
and~(\ref{sec: Relation between quantum RK Hamiltonians and stochastic classical systems})
we saw that the structure of the RK ground state is determined only by 
$
\beta\,E(s)
$,
while the energy spectrum is determined solely by the couplings
$
\omega_{s,s'}
$.~\cite{Henley-2004,Castelnovo-2005}
\item
If the couplings 
$
\omega_{s,s'}
$
are such that the energy spectrum has a non-degenerate gapped ground state 
(which is the only case considered in this work),
then Eq.~(\ref{eq: RK ground state 2})
parametrizes a large class of ground states, whereby 
$
\beta\,E(s)
$
controls the correlation length associated with the decay of correlation functions
such as 
Eq.~(\ref{eq: Correlation function of diagonal operators}).
\item
In the particular case $\beta=0$, the ground state 
Eq.~(\ref{eq: RK ground state 2})
acquires the form of a trivial product state. 
It will prove useful to work in this 
\lq\lq{infinite temperature}\rq\rq~
limit, where the correlation length is zero, in order to determine 
the form of the unitary transformation
Eq.~(\ref{eq: Definition of unitary mapping}).
\item
While the unitary mapping 
Eq.~(\ref{eq: Definition of unitary mapping})
leaves the energy spectrum unchanged,
it introduces a non-trivial entanglement structure in the 
SPT ground state 
Eq.~(\ref{eq: SPT ground state})
that is responsible for its universal topological properties.

\end{enumerate}

\section{SPT states in one dimension}
\label{sec: SPT states in one dimension}

In this section we derive parent Hamiltonians of one dimensional bosonic 
SPT states protected by 
$
\mathbb{Z}_{n}\times\mathbb{Z}_{n}
$
symmetry.
In a chain with periodic boundary conditions, each of these phases is described by a non-degenerate gapped symmetric ground state. In a chain with open boundary conditions, on the other hand, there remains a trivial phase with a non-degenerate ground state, while $n-1$ phases have $n$-fold degeneracy per edge, accounting for a total $n^2$-fold degeneracy of the ground state manifold in the thermodynamic limit.

Our aim is to construct the unitary transformations
Eq.~(\ref{eq: Definition of unitary mapping})
connecting the trivial ground state and the $n-1$ non-trivial SPT ground states. 
Our strategy in deriving such unitary mappings is to draw on the notion of entanglement spectrum degeneracy as follows.
As we pointed out in the remark (c) of 
Sec.~(\ref{sec: SPT-RK Hamiltonians}),
in the ``infinite temperature" limit $\beta = 0$
the ground state of the trivial SPT chain reduces to a 
product state.
The entanglement structure of the product state is as simple
as it gets, for the Schmidt decomposition with respect to any partition 
contains only one eigenvalue (equal to $1$).
We shall find the unitary transformation
Eq.~(\ref{eq: Definition of unitary mapping})
by demanding that, in the limit $\beta = 0$,
the entanglement spectrum of the non-trivial SPT 
ground state, for any partition of the chain, acquires 
an $n$-fold degeneracy.
Remarkably, we shall verify that these unitary mappings, via 
Eq.~(\ref{eq: Unitary mapping between Hamiltonians}), endow the parent Hamiltonians of non-trivial SPT chains with the required $n$-fold degeneracy of the energy spectrum per edge.
Once the unitary transformation
Eq.~(\ref{eq: Definition of unitary mapping})
is derived, we can obtain the most general form of the SPT
ground state 
Eq.~(\ref{eq: SPT ground state})
by allowing $\beta \neq 0$
without changing either the gapped nature of the many-body energy spectrum or
the topological properties of the ground state.
That this is true can be seen perturbatively: moving away from the $\beta=0$ limit
with $\beta << 1$ and $E_{G}(s)$ a local function 
in Eq.~(\ref{eq: Definition of the SPT Hamiltonian}), amounts to adding small local symmetry-preserving perturbations to the gapped $\beta=0$ theory, which therefore,
cannot immediately destroy the SPT phase.
Based on this we expect that, for $0< \beta \leq \beta_{c}$,
the ground state Eq.~(\ref{eq: SPT ground state}) describes a class of many-body SPT
ground states adiabatically connected to the $\beta=0$ limit.

\subsection
{
$\mathbb{Z}_{2}\,\times\,\mathbb{Z}_{2}$ SPT states in $\mathrm{d} = 1$
}
\label{sec: Z2 times Z2 in one dimension}

We consider a spin chain, 
with an even number $N_{\mathrm{s}}$ of sites,
divided into even and odd sublattices and
$
\mathbb{Z}_{2}\,\times\mathbb{Z}_{2}
$
symmetry generated by
\begin{equation}
\label{eq: Z2 times Z2 symmetry in one dimension}
\begin{split}
\widehat{S}^{(1)}_
{
\mathbb{Z}_{2}
}
=
\prod_{j \in \mathrm{even}}\,
X_{j}
\,,
\quad
\widehat{S}^{(2)}_
{
\mathbb{Z}_{2}
}
=
\prod_{j \in \mathrm{odd}}\,
X_{j}
\,.
\end{split}•
\end{equation}•
In the following,
$
\{ X, Y, Z\}
$
denote the three Pauli matrices.

We start by examining a trivial spin chain in the special limit where it is represented by a product state,
\begin{equation}
\label{eq: Trivial Z2 times Z2 ground state in one dimension}
\begin{split}
\ket{\Phi_{\mathbb{Z}_{2}\times\mathbb{Z}_{2}}}
=
\prod^{N_{\mathrm{s}}}_{j =1}
\,
\ket{\rightarrow}_{j}
=
\frac
{
1
}
{
2^{N_{\mathrm{s}}/2}
}
\sum_{ s }
\,
\ket{s} 
\,,
\end{split}
\end{equation}
where
$
X_{j}\,\ket{\rightarrow}_{j}
=
\ket{\rightarrow}_{j}
$
and $\{\,\ket{s}\,\}$ represents the $2^{N_{\mathrm{s}}}$ many-body spin states in the eigenbasis of $Z$ operators:
$
Z_{j}\,\ket{s_{j}}
=
s_{j}\,\ket{s_{j}}
$,
for
$
s_{j}
\pm
1
$.
We recognize 
Eq.~(\ref{eq: Trivial Z2 times Z2 ground state in one dimension}) as the 
ground state Eq.~(\ref{eq: RK ground state}) in the $\beta = 0$ limit.
It has the parent Hamiltonian
\begin{equation}
\label{eq: Trivial Hamiltonian for the Z2 times Z2 state in one dimension}
\begin{split}
H_{\mathbb{Z}_{2}\times\mathbb{Z}_{2}}
=
\omega_{0}
\,
\sum^{N_{\mathrm{s}}}_{j=1}\,
\left(
\,
1 - X_{j}
\,
\right)
\,,
\end{split}•
\end{equation}•
which is an RK Hamiltonian of the form Eq.~(\ref{eq: Definition of the RK Hamiltonian}) 
with $\beta = 0$
and
$
\omega_{s,s'}
=
\omega_{0}
>
0
$,
where the sum in Eq.~(\ref{eq: RK Hamiltonian}) extends
over pairs of states $(s,s')$ which differ by a single spin flip.

The product state form of the ground state
Eq.~(\ref{eq: Trivial Z2 times Z2 ground state in one dimension})
implies that its Schmidt decomposition, with respect to any partition $\Sigma_{i}$ between sites $i$ and $i+1$ of the lattice, contains a single Schmidt eigenvalue
(equal to $1$).

Now, for $\theta_{j} \in \mathbb{R}$, we introduce the unitary mapping
\begin{equation}
\label{eq: General form of the unitary mapping Z2 times Z2 in one dimension}
\begin{split}
\mathbb{W}^{\theta}_
{
\mathbb{Z}_{2}\times\mathbb{Z}_{2}
}
\equiv
\prod_{j}\,
e^
{
i\,\theta_{j}\,
\left(
1
-
Z_{j}\,Z_{j+1}
\right)/2
}
\,
\end{split}•
\end{equation}•
and the SPT state
\begin{equation}
\label{eq: Non-trivial Z2 times Z2 ground state in one dimension}
\begin{split}
\ket{\Psi_{\mathbb{Z}_{2}\times\mathbb{Z}_{2}}}
&\,
\equiv
\mathbb{W}^{\theta}_{\mathbb{Z}_{2}\times\mathbb{Z}_{2}}\,
\ket{\Phi_{\mathbb{Z}_{2}\times\mathbb{Z}_{2}}}
\\
&\,
\equiv
\frac
{
1
}
{
2^{N_{\mathrm{s}}/2}
}
\sum_{ s }
\,
e^
{
i\,W_{\mathbb{Z}_{2}\times\mathbb{Z}_{2}}(s)
}
\,
\ket{s} 
\\
&\,
=
\frac
{
1
}
{
2^{N_{\mathrm{s}}/2}
}
\sum_{ s }
\,
e^
{
i\,
\sum_{j}\,
\theta_{j}\,
\left(
1
-
s_{j}\,s_{j+1}
\right)/2
}
\ket{s} 
\,.
\end{split}
\end{equation}
The unitary transformation 
Eq.~(\ref{eq: General form of the unitary mapping Z2 times Z2 in one dimension}) 
endows the state 
Eq.~(\ref{eq: Non-trivial Z2 times Z2 ground state in one dimension}) 
with an amplitude $e^{i\,\theta_{j}}$ for every domain wall between neighbor spins in the many-body configuration $\ket{s}$.
We seek to find $\theta_{j}$ for which
Eq.~(\ref{eq: Non-trivial Z2 times Z2 ground state in one dimension}) 
describes the SPT ground state.

Due to the product state nature of 
Eq.~(\ref{eq: Trivial Z2 times Z2 ground state in one dimension}) 
and the pairwise entanglement induced by the unitary mapping 
Eq.~(\ref{eq: General form of the unitary mapping Z2 times Z2 in one dimension}),
one effortlessly finds that, for any partition $\Sigma_{i}$, the reduced density operator obtained by tracing over one of the subsystems is given by the $2 \times 2$ matrix

\begin{equation}
\rho_{i}
=
\frac{1}{2}\,
\begin{pmatrix}
1 & \cos{(\theta_{i})}
\\
\cos{(\theta_{i})} & 1
\end{pmatrix}•
\,.
\end{equation}•
For $\theta_{i}=0$ the above density matrix has a single non-zero Schmidt
eigenvalue.
The existence of $2$ degenerate Schmidt eigenvalues (equal to $1/2$) is verified for
\begin{equation}
\label{eq: Value of theta for the Z2 times Z2 in one dimension}
\theta_{i}
=
\pm \pi/2
\,.
\end{equation}•

Moreover, imposing that the unitary transformation
Eq.~(\ref{eq: General form of the unitary mapping Z2 times Z2 in one dimension})
commutes with either of the $\mathbb{Z}_{2}$ symmetries
in Eq.~(\ref{eq: Z2 times Z2 symmetry in one dimension}) yields the final form
\begin{equation}
\label{eq: Particular form of the unitary mapping Z2 times Z2 in one dimension}
\begin{split}
\mathbb{W}_
{
\mathbb{Z}_{2}\times\mathbb{Z}_{2}
}
=
\prod_{j}\,
e^
{
i\,\frac{\pi}{4}\,
\left(
1
-
Z_{2\,j-1}\,Z_{2\,j}
\right)
}
\,
e^
{
-i\,\frac{\pi}{4}\,
\left(
1
-
Z_{2\,j}\,Z_{2\,j+1}
\right)
}
\,.
\end{split}•
\end{equation}•

One then finds
\begin{equation}
\begin{split}
&\,
\mathbb{W}_{\mathbb{Z}_{2}\times\mathbb{Z}_{2}}
\,
X_{2j}
\,
\mathbb{W}_{\mathbb{Z}_{2}\times\mathbb{Z}_{2}}^{-1}
=
X_{2j}\,
Z_{2j-1}
\,
Z_{2j+1}
\,,
\\
&\,
\mathbb{W}_{\mathbb{Z}_{2}\times\mathbb{Z}_{2}}
\,
X_{2j+1}
\,
\mathbb{W}_{\mathbb{Z}_{2}\times\mathbb{Z}_{2}}^{-1}
=
X_{2j+1}\,
Z_{2j}
\,
Z_{2j+2}
\,.
\end{split}•
\end{equation}•

The operator
$
\mathbb{W}_{\mathbb{Z}_{2}\times\mathbb{Z}_{2}}
\,
X_{j}
\,
\mathbb{W}_{\mathbb{Z}_{2}\times\mathbb{Z}_{2}}^{-1}
$
can be regarded as a modified Pauli spin operator (since it is obtained
from $X_{j}$ by a unitary transformation) which is ``dressed" by the domain wall operator
$
Z_{j-1}\,Z_{j+1}
$
with support on the opposite sublattice.

So, under the unitary transformation 
Eq.~(\ref{eq: Particular form of the unitary mapping Z2 times Z2 in one dimension}), 
the SPT Hamiltonian at zero correlation length is
\begin{equation}
\label{eq: Non-trivial Z2 times Z2 Hamiltonian in one dimension}
\begin{split}
\mathcal{H}_{\mathbb{Z}_{2} \times \mathbb{Z}_{2}}
&\,
\equiv
\mathbb{W}_{\mathbb{Z}_{2} \times \mathbb{Z}_{2}}
\,
H_{\mathbb{Z}_{2} \times \mathbb{Z}_{2}}
\,
\mathbb{W}_{\mathbb{Z}_{2} \times \mathbb{Z}_{2}}^{-1}
\\
&\,=
\omega_{0}
\,
\sum^{N_{\mathrm{s}}}_{j=1}
\,
\Big(
1
-
Z_{j-1}
\,
X_{j}
\,
Z_{j+1}
\Big)
\,.
\end{split}•
\end{equation}•
We note that the model 
Eq.~(\ref{eq: Non-trivial Z2 times Z2 Hamiltonian in one dimension})
has been constructed in Ref.~\onlinecite{Chen-2014} using the concept of
decorated domain walls, while we have arrived on it by appealing to the notion
of entanglement spectrum via the unitary mapping 
Eq.~(\ref{eq: Particular form of the unitary mapping Z2 times Z2 in one dimension}).

Ground state degeneracy can be easily attested by studying this model
with open boundary conditions, where
\begin{equation}
\begin{split}
\mathcal{H}^{\textrm{open}}_{\mathbb{Z}_{2} \times \mathbb{Z}_{2}}
&\,=
\omega_{0}
\,
\sum^{N_{\mathrm{s}-1}}_{j=2}
\,
\Big(
1
-
Z_{j-1}
\,
X_{j}
\,
Z_{j+1}
\Big)
\,.
\end{split}•
\end{equation}•

The fact that the above Hamiltonian commutes with $Z_{1}$ and $Z_{N_{\mathrm{s}}}$
implies that there are $2$-fold degenerate states associated with the left and the right edges corresponding to states with
$
s_{1}
=
\pm
1
$
and
$
s_{N_{\mathrm{s}}}
=
\pm
1
$.
Thus the universal properties of the SPT state studied here
are encoded in the unitary mapping 
Eq.~(\ref{eq: Particular form of the unitary mapping Z2 times Z2 in one dimension}).
%

%
With the unitary transformation
Eq.~(\ref{eq: Particular form of the unitary mapping Z2 times Z2 in one dimension}),
we can obtain the more general form of the SPT
ground state 
Eq.~(\ref{eq: SPT ground state})
by allowing $\beta \neq 0$
without changing either the gapped nature of the many-body energy spectrum or
the topological properties of the ground state.
In this regard, the Hamiltonian 
Eq.~(\ref{eq: Non-trivial Z2 times Z2 Hamiltonian in one dimension}) is a particular example of a larger class of SPT models described in
Eq.~(\ref{eq: Definition of the SPT Hamiltonian}), with the phase factors
$
e^
{
i\,
W_{\mathbb{Z}_{2} \times \mathbb{Z}_{2}}(s)
}
$
given by acting with the unitary transformation
Eq.~(\ref{eq: Particular form of the unitary mapping Z2 times Z2 in one dimension}).
on the state $\ket{s}$.

\subsection
{
$\mathbb{Z}_{3}\,\times\,\mathbb{Z}_{3}$ SPT states in $ \mathrm{d} = 1$
}
\label{sec: Z3 times Z3 in one dimension}

We consider a spin chain, with an even number $N_{\mathrm{s}}$ of sites and
$
\mathbb{Z}_{3}\,\times\mathbb{Z}_{3}
$
symmetry generated by
\begin{equation}
\label{eq: Z3 times Z3 symmetry in one dimension}
\begin{split}
\widehat{S}^{(1)}_
{
\mathbb{Z}_{3}
}
=
\prod_{j \in \mathrm{even}}\,
\tau_{j}
\,,
\quad
\widehat{S}^{(2)}_
{
\mathbb{Z}_{3}
}
=
\prod_{j \in \mathrm{odd}}\,
\tau_{j}
\,.
\end{split}•
\end{equation}•
where, at each site $j$, we consider the operators
\begin{equation}
\begin{split}
\tau_{j}
=
\begin{pmatrix}
0 & 0 & 1
\\
1 & 0 & 0
\\
0 & 1 & 0
\end{pmatrix}•
\,,
\quad
\s_{j}
=
\begin{pmatrix}
1 & 0 & 0
\\
0 & \omega & 0
\\
0 & 0 & \omega^2
\end{pmatrix}•
\,,
\end{split}•
\end{equation}•
satisfying
$
\tau^{3}_{j}
=
\s^{3}_{j}
=
1
$,
$
\tau^{\dagger}_{j}
\,
\s_{j}
\,
\tau_{j}
=
\omega
\,
\s_{j}
$,
where
$\omega = e^{i\,2\pi/3}$.

Let $\ket{s_{i}}$,
for
$
s_{i}
=
0,1,2
$,
be the eigenstates of $\s_{i}$:
$
\s_{i}
\,
\ket{s_{i}}
=
\omega^{s_{i}}\,
\ket{s_{i}}
$.
As in 
Sec.~(\ref{sec: Z2 times Z2 in one dimension}), 
we start with a trivial $\mathbb{Z}_{3} \times \mathbb{Z}_{3}$
paramagnet described by the ground state
\begin{equation}
\label{eq: Trivial Z3 times Z3 ground state in one dimension}
\begin{split}
\ket{\Phi_{\mathbb{Z}_{3} \times \mathbb{Z}_{3}}}
\equiv
\prod_{j}\,
\Big(
\frac
{
\ket{0_{j}}
+
\ket{1_{j}}
+
\ket{2_{j}}
}
{
\sqrt{3}
}
\Big)
=
\frac
{
1
}
{
3^{N_{\mathrm{s}}/2}
}
\,
\sum_{\{ s \}}\,
\ket{s}
\,,
\end{split}•
\end{equation}•
where $\{\,\ket{s}\,\}$ represents the $3^{N_{\mathrm{s}}}$ many-body spin states in the eigenbasis of $\s$ operators.
We recognize Eq.~(\ref{eq: Trivial Z3 times Z3 ground state in one dimension}) as the 
RK ground state Eq.~(\ref{eq: RK ground state}) in the $\beta = 0$ limit.
It has the parent Hamiltonian
\begin{equation}
\label{eq: Trivial Hamiltonian for the Z3 times Z3 state in one dimension}
H_{\mathbb{Z}_{3} \times \mathbb{Z}_{3}}
=
\omega_{0}
\,
\sum^{N_{s}}_{j=1}\,
\left[
2
-
\left(
\tau_{j}
+
\tau^{\dagger}_{j}
\right)
\right]
\,,
\end{equation}•
which is an RK Hamiltonians of the form Eq.~(\ref{eq: Definition of the RK Hamiltonian}) 
with $\beta = 0$,
$
\omega_{s,s'}
=
\omega_{0}
>
0
$,
where the sum in 
Eq.~(\ref{eq: RK Hamiltonian})
extends over pairs of states $(s,s')$ which differ by a single $\mathbb{Z}_{3}$ spin flip.

We now consider, for $\theta_{j} \in \mathbb{R}$, the unitary transformation 
\begin{equation}
\label{eq: General form of the unitary mapping Z3 times Z3 in one dimension}
\begin{split}
\mathbb{W}^{\theta}_{\mathbb{Z}_{3} \times \mathbb{Z}_{3}}
=
\prod_{j}\,
\exp
{
\left[
i\,\theta_{j}
\sum^{2}_{a=1}\,
\frac
{
1-(\s^{\dagger}_{j}\,\s_{j+1})^{a}
}
{
(
\bar{\omega}^a - 1
)
\,
(
\omega^{a} - 1
)
}
\right]
}
\,
\end{split}•
\end{equation}•
and the SPT state
\begin{equation}
\label{eq: Non-trivial Z3 times Z3 ground state in one dimension}
\begin{split}
&\,
\ket{\Psi_{\mathbb{Z}_{3}\times\mathbb{Z}_{3}}}
\equiv
\mathbb{W}^{\theta}_{\mathbb{Z}_{3}\times\mathbb{Z}_{3}}\,
\ket{\Phi_{\mathbb{Z}_{3}\times\mathbb{Z}_{3}}}
\\
&\,
\equiv
\frac
{
1
}
{
3^{N_{\mathrm{s}}/2}
}
\sum_{ s }
\,
e^
{
i\,W_{\mathbb{Z}_{3}\times\mathbb{Z}_{3}}(s)
}
\,
\ket{s} 
\\
&\,
=
\frac
{
1
}
{
3^{N_{\mathrm{s}}/2}
}
\sum_{ s }
\,
\exp
{
\left[
\sum_{j}\,
i\,\theta_{j}
\sum^{2}_{a=1}\,
\frac
{
1
-
\omega^
{
a\,\left(
s_{j+1}
-
s_{j}
\right)
}
}
{
(
\bar{\omega}^a - 1
)
\,
(
\omega^{a} - 1
)
}
\right]
}
\ket{s} 
\,.
\end{split}
\end{equation}

The unitary transformation
Eq.~(\ref{eq: General form of the unitary mapping Z3 times Z3 in one dimension}) endows the state 
Eq.~(\ref{eq: Non-trivial Z3 times Z3 ground state in one dimension}) 
with an amplitude $e^{i\,\theta_{j}}$ for every 
pair of neighbor spins $j$ and $j+1$ for which 
$
s_{j}
\neq
s_{j+1}
~
(\textrm{mod}~3)
$
in the many-body configuration $\ket{s}$.

Due to the product state nature of 
Eq.~(\ref{eq: Trivial Z3 times Z3 ground state in one dimension}) 
and the pairwise entanglement induced by the unitary mapping 
Eq.~(\ref{eq: General form of the unitary mapping Z3 times Z3 in one dimension}),
one effortlessly verifies that, for any partition $\Sigma_{i}$, the reduced density operator obtained by tracing over one of the subsystems is given by the $3 \times 3$ matrix
\begin{equation}
\rho_{i}
=
\frac{1}{3}
\,
\left(
\begin{array}{ccc}
 1 & f(\theta_{i}) & f(\theta_{i}) \\
 f(\theta_{i}) & 1 & f(\theta_{i}) \\
 f(\theta_{i}) & f(\theta_{i}) & 1 \\
\end{array}
\right)
\,,
\end{equation}•
where
$
f(\theta)
=
\frac{1}{3} (1+2 \cos{(\theta)} )
$.
For $\theta_{i}=0$ the above density matrix has a single non-zero Schmidt
eigenvalue.
The existence of $3$ degenerate Schmidt eigenvalues (equal to $1/3$) is 
then verified for
\begin{equation}
\theta^{(p)}_{i}
=
\frac{2\pi}{3}\,p
\,,
\quad
p = 1, 2
\,.
\end{equation}•

Moreover, imposing that the unitary transformation 
Eq.~(\ref{eq: General form of the unitary mapping Z3 times Z3 in one dimension})
commutes with either of the $\mathbb{Z}_{3}$ symmetries
in Eq.~(\ref{eq: Z3 times Z3 symmetry in one dimension}) yields the final form
\begin{widetext}
\begin{equation}
\label{eq: Particular form of the unitary mapping Z3 times Z3 in one dimension}
\begin{split}
\mathbb{W}_{\mathbb{Z}_{3}\times\mathbb{Z}_{3}}^{(p)}
&\,=
\prod_{j}\,
\exp
{
\Big[
i\,\frac{2 \pi p}{3}
\,
\sum^{3}_{a=1}\,
\frac
{
1
-
(
\s^{\dagger}_{2j-1}\,
\s_{2j}
)^a
}
{
(
\omega^a - 1
)
\,
(
\bar{\omega}^a - 1
)
}
\Big]
}
\,
\exp
{
\Big[
-i\,\frac{2 \pi p}{3}
\,
\sum^{2}_{a=1}\,
\frac
{
1
-
(
\s^{\dagger}_{2j}\,
\s_{2j+1}
)^a
}
{
(
\omega^a - 1
)
\,
(
\bar{\omega}^a - 1
)
}
\Big]
}
\,,
\quad
p = 1,2
\,.
\end{split}•
\end{equation}•
\end{widetext}
Eq.~(\ref{eq: Particular form of the unitary mapping Z3 times Z3 in one dimension})
establishes two unitary mappings between the trivial SPT ground state and the two non-trivial SPT ground states.

Moreover we find that 
\begin{equation}
\begin{split}
&\,
\mathbb{W}_{\mathbb{Z}_{3}\times\mathbb{Z}_{3}}^{(p)}
\,
\tau_{2j}
\,
\left(
\mathbb{W}_{\mathbb{Z}_{3}\times\mathbb{Z}_{3}}^{(p)}
\right)^{-1}
=
\tau_{2j}
\,
(
\s_{2j-1}\,\s^{\dagger}_{2j+1}
)^{p}
\,,
\\
&\,
\mathbb{W}_{\mathbb{Z}_{3}\times\mathbb{Z}_{3}}^{(p)}
\,
\tau_{2j+1}
\,
\left(
\mathbb{W}_{\mathbb{Z}_{3}\times\mathbb{Z}_{3}}^{(p)}
\right)^{-1}
=
\tau_{2j+1}
\,
(
\s^{\dagger}_{2j}\,
\s_{2j+2}
)^{p}
\,,
\end{split}•
\end{equation}•
so that the operator $\tau_{j}$ gets ``dressed" by a domain wall like operator --
carrying $\mathbb{Z}_{3}$ charge $p = 1,2$ -- with support on the opposite sublattice.

The SPT Hamiltonians in the $\beta=0$ limit then read
\begin{widetext}
\begin{equation}
\label{eq: Non-trivial Z3 times Z3 Hamiltonian in one dimension}
\begin{split}
\mathcal{H}^{(p)}_{\mathbb{Z}_{3} \times \mathbb{Z}_{3}}
&\,=
\mathbb{W}_{\mathbb{Z}_{3}\times\mathbb{Z}_{3}}^{(p)}
\,
H_{\mathbb{Z}_{3} \times \mathbb{Z}_{3}}
\,
\left(
\mathbb{W}_{\mathbb{Z}_{3}\times\mathbb{Z}_{3}}^{(p)}
\right)^{-1}
\\
&\,
=
\omega_{0}
\,
\sum_{j}\,
\Big\{
\left[
\,
1
-
\tau_{2j}
\,
(
\s_{2j-1}\,
\s^{\dagger}_{2j+1}
)^{p}
\,
\right]
+
\textrm{H.c.}
\Big\}
+
\omega_{0}
\,
\sum_{j}\,
\Big\{
\left[
\,
1
-
\tau_{2j+1}
\,
(
\s^{\dagger}_{2j}\,
\s_{2j+2}
)^{p}
\,
\right]
+
\textrm{H.c.}
\Big\}
\,,
\end{split}
\end{equation}
\end{widetext}
for 
$
p = 1,2.
$

As in Sec.~(\ref{sec: Z2 times Z2 in one dimension}), 
degeneracy of the ground state energy manifold can be checked by placing this system in an open chain, in which case, one finds that the SPT Hamiltonians commutes with $\s_{1}$ and $\s_{N_{\mathrm{s}}}$, thus implying a $3$-fold degenerate state associated to the left and right edges.
Thus the universal properties of the SPT state studied here
are encoded in the unitary mapping 
Eq.~(\ref{eq: Particular form of the unitary mapping Z3 times Z3 in one dimension}).

With the unitary transformation
Eq.~(\ref{eq: Particular form of the unitary mapping Z3 times Z3 in one dimension})
we can obtain the more general form of the SPT ground state 
Eq.~(\ref{eq: SPT ground state})
by allowing $\beta \neq 0$ without changing either the gapped nature of the many-body energy spectrum or the topological properties of the ground state.
In this regard, the Hamiltonian 
Eq.~(\ref{eq: Non-trivial Z3 times Z3 Hamiltonian in one dimension}) provides a particular example of a larger class of models described in
Eq.~(\ref{eq: Definition of the SPT Hamiltonian}), with the phase factors
$
e^
{
i\,
W_{\mathbb{Z}_{3} \times \mathbb{Z}_{3}}(s)
}
$
given by acting with the unitary transformation
Eq.~(\ref{eq: Particular form of the unitary mapping Z3 times Z3 in one dimension}).
on the state $\ket{s}$.

\subsection
{
$\mathbb{Z}_{n}\,\times\,\mathbb{Z}_{n}$ SPT states in $ \mathrm{d} = 1$
}
\label{sec: Zn times Zn in one dimension}

We generalize the findings of Secs.~(\ref{sec: Z2 times Z2 in one dimension}) 
and~(\ref{sec: Z3 times Z3 in one dimension}) 
to $\mathbb{Z}_{n}\,\times\,\mathbb{Z}_{n}$
SPT states (see the Appendix for definitions and useful formulas).
We arrive at $n-1$
$\mathbb{Z}_{n}\,\times\,\mathbb{Z}_{n}$
symmetric unitary transformations, labeled by
$
p = 1, ... , n-1
$,
\begin{widetext}
\begin{equation}
\label{eq: Particular form of the unitary mapping Zn times Zn in one dimension}
\begin{split}
\mathbb{W}_{\mathbb{Z}_{n}\,\times\,\mathbb{Z}_{n}}^{(p)}
&\,=
\prod_{j}\,
\exp
{
\Big[
i\,\frac{2 \pi p}{n}
\,
\sum^{n-1}_{a=1}\,
\frac
{
1
-
(
\s^{\dagger}_{2j-1}\,
\s_{2j}
)^a
}
{
(
\omega^a - 1
)
\,
(
\bar{\omega}^a - 1
)
}
\Big]
}
\,
\exp
{
\Big[
-i\,\frac{2 \pi p}{n}
\,
\sum^{n-1}_{a=1}\,
\frac
{
1
-
(
\s^{\dagger}_{2j}\,
\s_{2j+1}
)^a
}
{
(
\omega^a - 1
)
\,
(
\bar{\omega}^a - 1
)
}
\Big]
}
\,.
\end{split}•
\end{equation}•
\end{widetext}

It can be shown, using
Eqs.~(\ref{eq: Zn algebra appendix}), 
~(\ref{eq: Algebra Zn operators})
and~(\ref{eq: Identity Zn operators}), that
 \begin{equation}
\begin{split}
&\,
\mathbb{W}_{\mathbb{Z}_{n}\,\times\,\mathbb{Z}_{n}}^{(p)}
\,
\tau_{2j}
\,
\left(
\mathbb{W}_{\mathbb{Z}_{n}\,\times\,\mathbb{Z}_{n}}^{(p)}
\right)^{-1}
=
\tau_{2j}
\,
(
\s_{2j-1}\,\s^{\dagger}_{2j+1}
)^{p}
\,,
\\
&\,
\mathbb{W}_{\mathbb{Z}_{n}\,\times\,\mathbb{Z}_{n}}^{(p)}
\,
\tau_{2j+1}
\,
\left(
\mathbb{W}_{\mathbb{Z}_{n}\,\times\,\mathbb{Z}_{n}}^{(p)}
\right)^{-1}
=
\tau_{2j+1}
\,
(
\s^{\dagger}_{2j}\,
\s_{2j+2}
)^{p}
\,,
\end{split}•
\end{equation}•
for 
$
p = 1, ... , n-1
$.

The 
$
\mathbb{Z}_{n}\,\times\,\mathbb{Z}_{n}
$
SPT Hamiltonians in the $\beta = 0$ limit are of the same form as 
Eq.~(\ref{eq: Non-trivial Z3 times Z3 Hamiltonian in one dimension})
with
$
p = 1, ..., n-1
$.
We notice that this class of one dimensional SPT models has been studied in 
Ref.~\onlinecite{Geraedts-2014} without reference to the connection
between unitary transformations and ground state entanglement spectrum that we 
are exploring here.
Moreover, in our formalism, we do not need to be restricted to the $\beta=0$
limit for, with the unitary transformation 
Eq.~(\ref{eq: Particular form of the unitary mapping Zn times Zn in one dimension}),
we can construct a large class of SPT-RK Hamiltonians of the form  
Eq.~(\ref{eq: Definition of the SPT Hamiltonian})
with the phase factors
$
e^
{
i\,
W_{\mathbb{Z}_{n} \times \mathbb{Z}_{n}}(s)
}
$
given by acting with the unitary transformation
Eq.~(\ref{eq: Particular form of the unitary mapping Zn times Zn in one dimension})
on the basis state $\ket{s}$.

\section{
SPT state in one dimension with time-reversal symmetry
}
\label{sec: Z2T SPT state in one dimension}

We now study the one dimensional bosonic SPT states protected by time-reversal symmetry which, being an anti-unitary symmetry, requires conditions
Eq.~(\ref{eq: Transformation of E and W under G anti-unitary}) to be satisfied.

The action of time-reversal symmetry shall be represented by the anti-unitary operator
\begin{subequations}
\label{eq: Z2T symmetry in one dimension}
$
\Theta
$,
\begin{equation}
\Theta\,\Theta^{-1}
=
1
\,,
\quad
\Theta^2 = 1
\,,
\end{equation}•
where we work with the representation
\begin{equation}
\begin{split}
\Theta
&\,=
\Big(
\,
\prod_{j}\,X_{j}
\,
\Big)
K
\,,
\end{split}•
\end{equation}•
\end{subequations}•
with $K$ denoting the complex conjugation operator.

A one dimensional chain of $N_{\mathrm{s}}$ sites
in the trivial 
$
\mathbb{Z}^{T}_{2}
$
insulating phase can be described by the Hamiltonian
\begin{equation}
\label{eq: Trivial Hamiltonian for the Z2T state in one dimension}
\begin{split}
H_{\mathbb{Z}^{T}_{2}}
&\,=
\omega_{0}\,
\sum^{N_{\mathrm{s}}}_{j=1}\,
\left(
1 - X_{j}
\right)
\,,
\end{split}•
\end{equation}•
which has the time-reversal symmetric ground state
\begin{equation}
\label{eq: Trivial Z2T ground state in one dimension}
\begin{split}
\ket{\Phi_{\mathbb{Z}^{T}_{2}}}
=
\frac
{
1
}
{
2^{N_{\mathrm{s}}/2}
}
\,
\sum_{ s }
\,
\ket
{s}
\,,
\end{split}•
\end{equation}•
where
$\{\,\ket{s}\,\}$ represents the $2^{N_{\mathrm{s}}}$ many-body spin states in the eigenbasis of $Z$ operators:
$
Z_{j}\,\ket{s_{j}}
=
s_{j}\,\ket{s_{j}}
$,
for
$
s_{j}
=
\pm
1
$.
We recognize 
Eq.~(\ref{eq: Trivial Z2T ground state in one dimension}) 
as the 
ground state Eq.~(\ref{eq: RK ground state}) in the $\beta = 0$ limit
and
Eq.~(\ref{eq: Trivial Hamiltonian for the Z2T state in one dimension})
as an RK Hamiltonian of the form Eq.~(\ref{eq: Definition of the RK Hamiltonian}) 
with $\beta = 0$
and
$
\omega_{s,s'}
=
\omega_{0}
>
0
$,
where the sum in Eq.~(\ref{eq: RK Hamiltonian}) extends
over pairs of states $(s,s')$ which differ by a single spin flip.

We now obtain the unitary mapping $\mathbb{W}_{\mathbb{Z}^{T}_{2}}$, which,
applied on the trivial ground state 
Eq.~(\ref{eq: Trivial Z2T ground state in one dimension}),
yields the non-trivial SPT state 
\begin{equation}
\label{eq: Non-trivial Z2T ground state in one dimension}
\begin{split}
\ket{\Psi_{\mathbb{Z}^{T}_{2}}}
\equiv
\mathbb{W}_{\mathbb{Z}^{T}_{2}}\,
\ket{\Phi_{\mathbb{Z}^{T}_{2}}}
\equiv
\frac
{
1
}
{
2^{N_{\mathrm{s}}/2}
}
\,
\sum_{ s }
\,
e^
{
i\,W_{\mathbb{Z}^{T}_{2}}(s)
}
\,
\ket
{s}
\,,
\end{split}•
\end{equation}•
wherein the action of $\mathbb{W}_{\mathbb{Z}^{T}_{2}}$ 
on the many-body basis
$\{ \ket{s} \}$ is defined as
\begin{equation}
\label{eq: Action of W for the Z2T state on the basis elements}
\begin{split}
\mathbb{W}_{\mathbb{Z}^{T}_{2}}\,\ket{s}
=
e^
{
i\,W_{\mathbb{Z}^{T}_{2}}(s)
}
\,
\ket
{s}
\,.
\end{split}•
\end{equation}•

Invariance of the SPT ground state
$
\ket{\Psi_{\mathbb{Z}^{T}_{2}}}
$
under time-reversal symmetry
Eq.~(\ref{eq: Z2T symmetry in one dimension}), 
according to 
Eq.~(\ref{eq: Transformation of E and W under G anti-unitary}),
implies
\begin{equation}
\label{eq: Z2T condition on W}
\begin{split}
W_{\mathbb{Z}^{T}_{2}}(s)
=
-
W_{\mathbb{Z}^{T}_{2}}(-s)
+
2 \pi \times \textrm{integer}
\,.
\end{split}•
\end{equation}•

In light of the discussion in Sec.~(\ref{sec: SPT states in one dimension}),
if we assume for $\mathbb{W}_{\mathbb{Z}^{T}_{2}}$ an ansatz that leads to the entanglement of nearest neighbor spins, than
the only transformation (up to a trivial global phase) consistent with 
Eq.~(\ref{eq: Z2T condition on W}) has
\begin{equation}
\label{eq: W_{G}(s) for the Z2T state in one dimension}
W_{\mathbb{Z}^{T}_{2}}(s)
=
\frac{\pi}{4}\,
\sum_{j}
\,
\left(
1 - s_{j}\,s_{j+1}
\right)
\,
\end{equation}•
due to the fact that, with periodic boundary conditions,
$
\sum_{j}\,
(1 - s_{j}\,s_{j+1})/2
$
is an even number. 
Moreover, comparing 
Eqs.~(\ref{eq: Action of W for the Z2T state on the basis elements})
and~(\ref{eq: W_{G}(s) for the Z2T state in one dimension})
with the results obtained in Sec~(\ref{sec: Z2 times Z2 in one dimension}), we conclude that, for any partition $\Sigma_{i}$ between sites $i$ and $i+1$, there are $2$ degenerate (equal to $1/2$) Schmidt eigenvalues and that 
the SPT parent Hamiltonian for the ground state
Eq.~(\ref{eq: Non-trivial Z2T ground state in one dimension})
reads
\begin{equation}
\label{eq: Non-trivial Z2T Hamiltonian in one dimension}
\begin{split}
\mathcal{H}_{\mathbb{Z}^{T}_{2}}
\equiv
\mathbb{W}_{\mathbb{Z}^{T}_{2}}
\,
H_{\mathbb{Z}^{T}_{2}}
\,
\mathbb{W}_{\mathbb{Z}^{T}_{2}}^{-1}
&\,=
\omega_{0}\,
\sum^{N_{\mathrm{s}}}_{j=1}
\,
\left(
1
+
Z_{j-1}
\,
X_{j}
\,
Z_{j+1}
\right)
\,.
\end{split}•
\end{equation}•

Since this is essentially the same model as the 
$
\mathbb{Z}_{2}\times\mathbb{Z}_{2}
$
spin chain discussed in Sec~(\ref{sec: Z2 times Z2 in one dimension}),
we conclude that the
$
\mathbb{Z}_{2}^{T}
$
symmetric SPT Hamiltonian has $2$-fold degeneracy of the ground state per edge.

With the unitary transformation
Eqs.~(\ref{eq: Action of W for the Z2T state on the basis elements})
and~(\ref{eq: W_{G}(s) for the Z2T state in one dimension})
we can obtain the most general form of the SPT ground state 
Eq.~(\ref{eq: SPT ground state})
by allowing $\beta \neq 0$ without changing either the gapped nature of the many-body energy spectrum or the topological properties of the ground state.
In this regard, the Hamiltonian 
Eq.~(\ref{eq: Non-trivial Z2T Hamiltonian in one dimension}) 
provides a particular example of a larger class of models described in
Eq.~(\ref{eq: Definition of the SPT Hamiltonian}), with the phase factors
$
e^
{
i\,
W_{\mathbb{Z}_{2}^{T}}(s)
}
$
given in
Eq.~(\ref{eq: W_{G}(s) for the Z2T state in one dimension}).

\section{SPT states in two dimensions}
\label{SPT states in two dimensions}

\subsection{$\mathbb{Z}_{2}$ SPT states in $d=2$}
\label{subsec:Z2 SPT states in d=2}

The classification of bosonic SPT states in two dimensions, protected by 
$\mathbb{Z}_{n}$ symmetry, asserts the existence of $n$ gapped phases of matter.~\cite{Chen-2012-a,Chen-2013,Lu-2012}
From the point of view of the one dimensional edge, these $n$ phases can be distinguished by the way right- and left-moving degrees of freedom transform under the symmetry: while in the trivial phase right- and left-moving modes carry the same $\mathbb{Z}_{n}$ charges, their $\mathbb{Z}_{n}$ charges are different for each of the other $n-1$ non-trivial SPT states.~\cite{Lu-2012,Chen-2012-b,Santos-2014}
As a consequence, in each of these $n-1$ non-trivial SPT states
the edge states cannot be gapped and symmetry preserving at the same time.

Levin and Gu have constructed a model on a triangular lattice 
that describes a $\mathbb{Z}_{2}$ SPT paramagnet in two dimensions.~\cite{Levin-2012}
The $\mathbb{Z}_{2}$ symmetry is generated by
\begin{equation}
\label{eq: Z2 symmetry in 2 dimensions}
\widehat{S}_{\mathbb{Z}_{2}} = \prod_{j}\,X_{j}
\,,
\end{equation}•
where
$
\{ X, Y, Z\}
$
denote the three Pauli matrices
and
$
Z_{j}\,\ket{s_{j}}
=
s_{j}\,\ket{s_{j}}
$,
for
$s_{j}=\pm 1$.

In Ref.~\onlinecite{Levin-2012} the two kinds of paramagnetic ground states are represented by
\begin{equation}
\label{eq: Trivial Z2 paramagnet}
\begin{split}
\ket{\Phi_{\mathbb{Z}_{2}}}
=
\frac{1}{2^{N_{\textrm{s}}/2}}\,
\sum_{ s }\,
\ket{s}
\,,
\end{split}•
\end{equation}•
for the trivial paramagnet, and
\begin{equation}
\label{eq: Non-trivial Z2 paramagnet}
\begin{split}
\ket{\Psi_{\mathbb{Z}_{2}}}
=
\frac{1}{2^{N_{\textrm{s}}/2}}\,
\sum_{ s }\,
e^
{
i\,\pi\,L(s)
}\,
\ket{s}
\,,
\end{split}•
\end{equation}•
for the non-trivial paramagnet.
$N_{\textrm{s}}$ denotes the number of sites of the triangular lattice and $L(s) \in \mathbb{N}$ counts the number of loops that separate 
domain-wall regions 
for each one of the 
$
2^{N_{\textrm{s}}}
$
many-body spin configurations $\ket{s}$
(see Figs.~\ref{fig:1},~\ref{fig:2}, and~\ref{fig:3}).
Thus while the coefficients in the expansion Eq.~(\ref{eq: Trivial Z2 paramagnet}) all have the same sign, in the non-trivial SPT ground state 
Eq.~(\ref{eq: Non-trivial Z2 paramagnet}), the sign of the coefficients depends on the number of loops $L(s)$. This non-trivial (real) phase factor was shown in 
Ref.~\onlinecite{Levin-2012} to 
be responsible for the universal properties of the $\mathbb{Z}_{2}$ non-trivial paramagnetic state.

From 
Eqs.~(\ref{eq: Trivial Z2 paramagnet}) 
and~(\ref{eq: Non-trivial Z2 paramagnet}),
we identify the ground states 
of Ref.~\onlinecite{Levin-2012}
with Eqs.~(\ref{eq: RK ground state})
and~(\ref{eq: SPT ground state})
in the limit $\beta=0$,
and the unitary mapping 
$\mathbb{W}_{\mathbb{Z}_{2}}$ between the two classes of paramagnets,
defined by its action on the basis $\{ \ket{s} \}$, to be 
\begin{equation}
\label{eq: Phase factor in the Z2 SPT state in 2 dimensions}
\mathbb{W}_{\mathbb{Z}_{2}}\,\ket{s}
=
e^
{
i\,W_{\mathbb{Z}_{2}}(s)
}
\,\ket{s}
=
e^
{
i\,\pi\,L(s)
}
\,\ket{s}
\,.
\end{equation}•

Given that the unitary transformation defined in 
Eq.~(\ref{eq: Phase factor in the Z2 SPT state in 2 dimensions})
encodes the universal properties of the two dimensional 
$\mathbb{Z}_{2}$ SPT paramagnet, we can
move away from the ``infinite temperature" limit $\beta = 0$ 
studied in Ref.~(\onlinecite{Levin-2012})
by allowing a larger class of
$
\mathbb{Z}_{2}
$
SPT paramagnets with ground state given 
by
Eq.~(\ref{eq: SPT ground state})
and 
microscopic Hamiltonians
given by
Eq.~(\ref{eq: Definition of the SPT Hamiltonian}).

\subsection{$\mathbb{Z}_{2} \times \mathbb{Z}_{2}$ SPT states in $d=2$}
\label{subsec: Z2 times Z2 SPT states in d=2}


\begin{figure}[h!]
\includegraphics[width=0.3\textwidth]{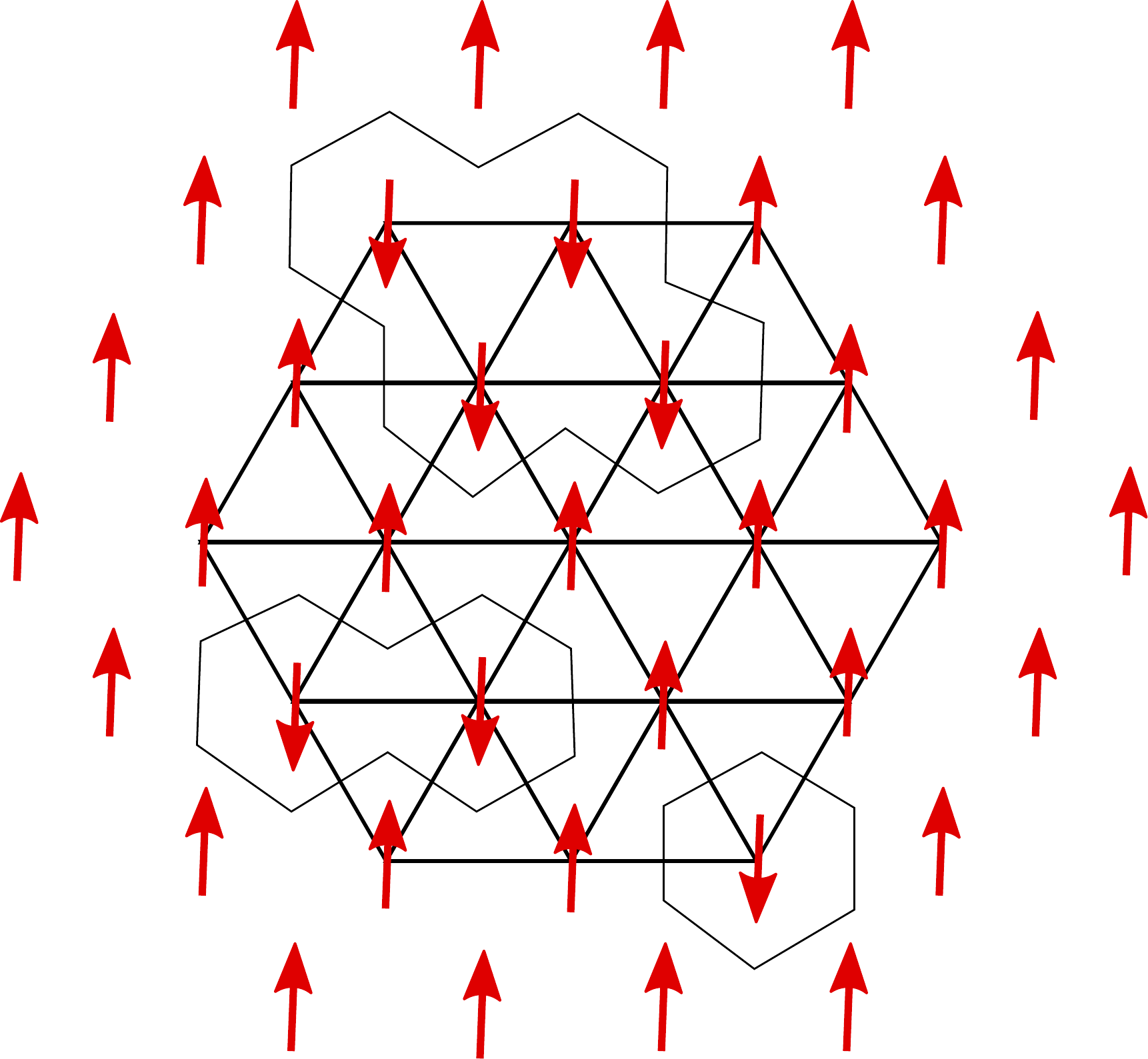}
\caption
{
(Color online) A particular many-body configuration $\ket{s^{(1)}}$ of spins 
$
Z^{(1)}_{j}
$ 
(red),
where the loops are defined along the domain walls and have support on the 
dual lattice.
As in Ref.~\onlinecite{Levin-2012}, in order to determine the 
projection of the $\mathbb{Z}_{2}$ symmetry on the edge, 
we take the outer spins in a reference state where they are all pointing
up.
}
\label{fig:1}
\end{figure}

\begin{figure}[h!]
\includegraphics[width=0.3\textwidth]{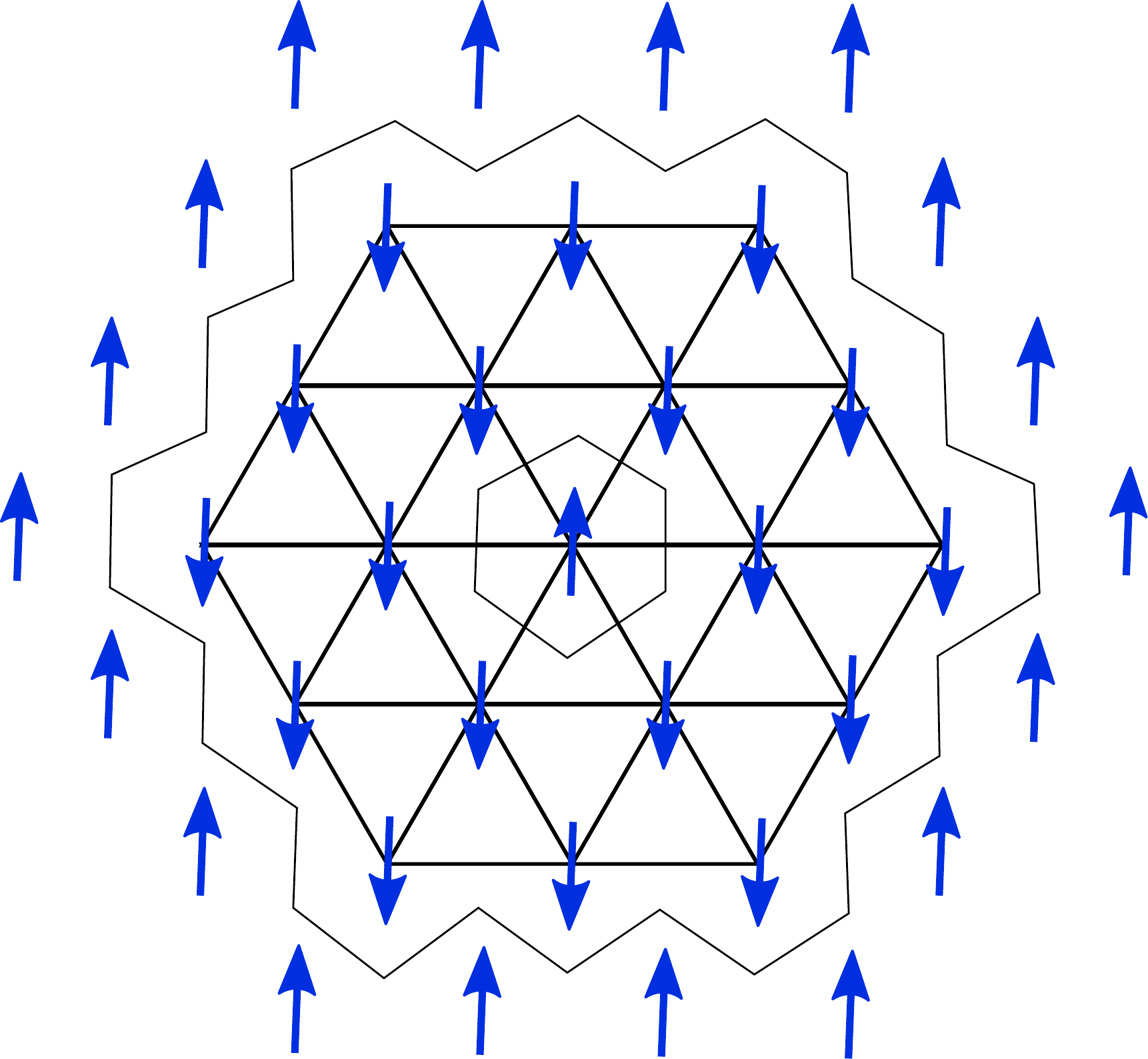}
\caption
{
(Color online)
A particular many-body configuration $\ket{s^{(2)}}$ of spins 
$
Z^{(2)}_{j}
$ 
(blue).
}
\label{fig:2}
\end{figure}

\begin{figure}[h!]
\includegraphics[width=0.3\textwidth]{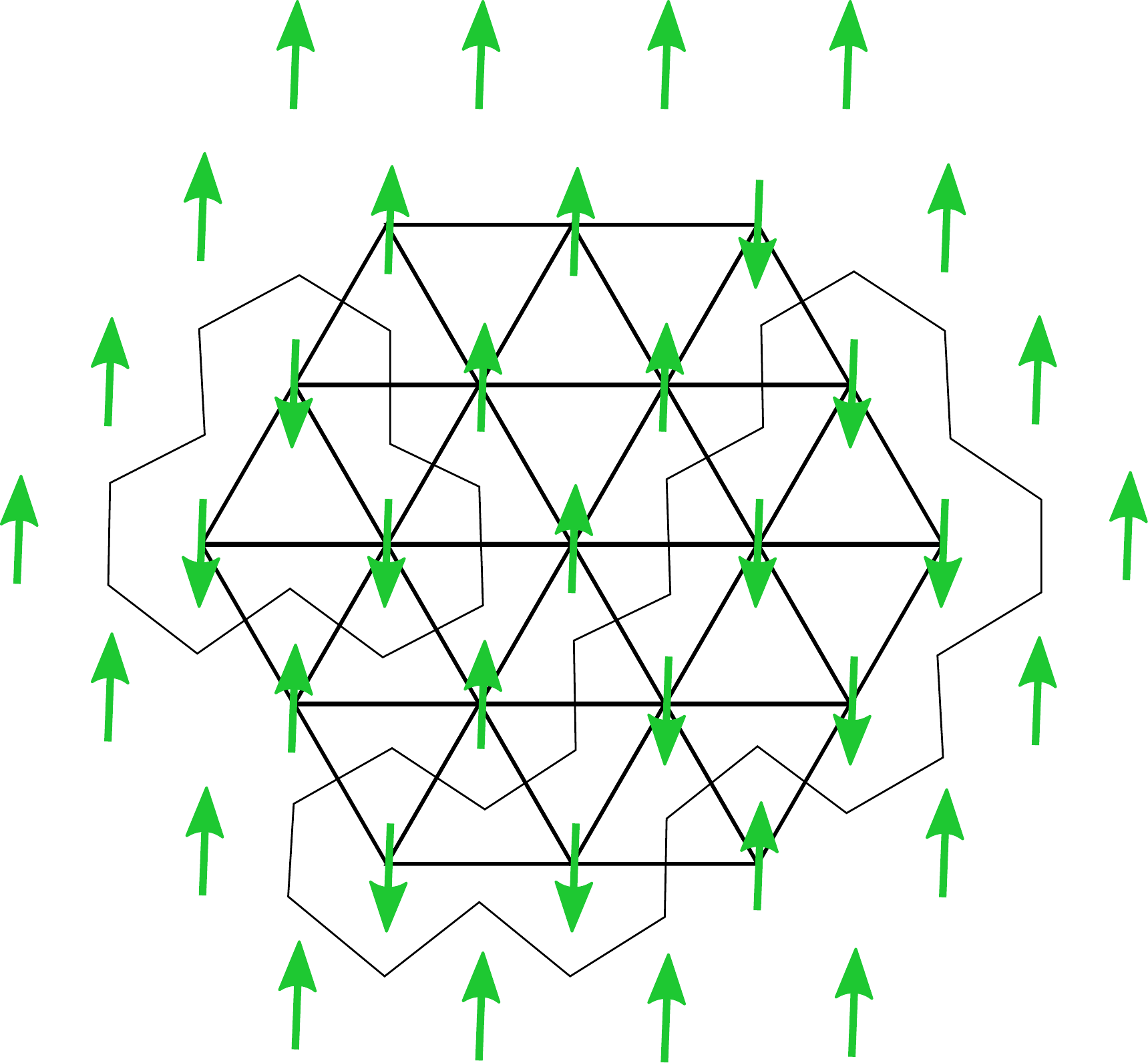}
\caption
{
(Color online)
A particular many-body configuration $\ket{s^{(12)}}$ of the spins
$
Z^{(12)}_{j}
=
Z^{(1)}_{j}
\,
Z^{(2)}_{j}
$
(green).
Note how the direction of the arrows at every site is consistent
with 
Figs.~\ref{fig:1} and~\ref{fig:2}.
}
\label{fig:3}
\end{figure}


In Ref.~\onlinecite{J-Wang-2014}, the classification of two dimensional bosonic SPT states with 
$\mathbb{Z}_{n} \times \mathbb{Z}_{m}$ 
symmetry was studied from the point of view of the effective one dimensional edge degrees of freedom.
The non-trivial SPT edge states were there shown 
to be associated with non-trivial $3$-cocycles in the group cohomology.
However, no microscopic theory for the two-dimensional bulk states has been 
investigated. 
In this section, we construct a microscopic theory for the 
$\mathbb{Z}_{2} \times \mathbb{Z}_{2}$ bosonic SPT states using the formalism of RK Hamiltonians and show how the topological properties of the ground state give rise to
the edge physics studied in 
Ref.~\onlinecite{J-Wang-2014}.

In addressing SPT states with $\mathbb{Z}_{2} \times \mathbb{Z}_{2}$ symmetry, 
we allow, at each site $j$ of the lattice, two spin species represented by Pauli operators 
$
\{\,X^{(1)}_{j}, Y^{(1)}_{j}, Z^{(1)}_{j}\,\}
$
and
$
\{\,X^{(2)}_{j}, Y^{(2)}_{j}, Z^{(2)}_{j}\,\}
$,
where the 
$\mathbb{Z}_{2} \times \mathbb{Z}_{2}$ symmetry is generated by
\begin{equation}
\label{eq: Z2 times Z2 symmetry}
\begin{split}
\widehat{S}^{(1)}_{\mathbb{Z}_{2}}
=
\prod_{j}\,X^{(1)}_{j}
\,,
\quad
\widehat{S}^{(2)}_{\mathbb{Z}_{2}}
=
\prod_{j}\,X^{(2)}_{j}
\,.
\end{split}•
\end{equation}•

We immediately realize that there should exist at least four kinds of paramagnets, which are obtained by stacking two decoupled $\mathbb{Z}_{2}$ symmetric systems.
These four states of matter are parametrized by two integer numbers, 
$p_{1}, p_{2} \in \{ 0, 1 \}$,
which contribute to the expansion of the ground state with the (real) phase factor
$
\exp
{
\left\{
i\,\pi\,
\left[
p_{1}\,L(s^{(1)})
+
p_{2}\,L(s^{(2)})
\right]
\right\}
}
$,
where 
$L(s^{(1)})$ and $L(s^{(2)})$ count, respectively, the number of loops 
defined by domain wall configurations formed by spins of species $1$ and $2$,
as in Figs.~\ref{fig:1} and~\ref{fig:2}.

In order to obtain $\mathbb{Z}_{2} \times \mathbb{Z}_{2}$ symmetric states that go beyond the tensor product of two $\mathbb{Z}_{2}$ symmetric states, we realize that, at each site $j$, we can consider an independent Ising spin
$
Z^{(12)}_{j}
\equiv
Z^{(1)}_{j}\,Z^{(2)}_{j}
$
with eigenvalues
$
s^{(12)}_{j}
=
s^{(1)}_{j}
\,
s^{(2)}_{j}
$.

We then propose 
\begin{widetext}
\begin{equation}
\label{eq: Z2 times Z2 paramagnet ground state}
\begin{split}
\ket
{
\Psi^
{
2 \mathrm{d}
}
_
{
\mathbb{Z}_{2}\times\mathbb{Z}_{2}
}
}
=
\frac{1}{2^{N_{\mathrm{s}}}}
\,
\sum_{ s^{(1)} }\,
\sum_{ s^{(2)} }\,
e^
{
i\,\pi\,
\left[
p_{1}\,L(s^{(1)})
+
p_{2}\,L(s^{(2)})
+
p_{12}\,L(s^{(12)})
\right]
}
\,
\ket
{
s^{(1)}; s^{(2)}
}
\,
\end{split}•
\end{equation}•
\end{widetext}
as a description of the $8$ classes of
$
\mathbb{Z}_{2} \times \mathbb{Z}_{2}
$
symmetric ground states parametrized by $3$ binary indices 
$
p_{1}, p_{2}, p_{12} \in \{ 0, 1 \}
$.
We adopt the notation
$
\ket{s^{(1)}}
\otimes
\ket{s^{(2)}}
\equiv
\ket{s^{(1)};s^{(2)}}
$.
When 
$
p_{1} = p_{2} = p_{12} = 0
$,
Eq.~(\ref{eq: Z2 times Z2 paramagnet ground state}) describes the trivial ground state
\begin{equation}
\label{eq: Trivial Z2 times Z2 paramagnet ground state}
\begin{split}
\ket
{
\Phi^
{
2 \mathrm{d}
}
_
{
\mathbb{Z}_{2}\times\mathbb{Z}_{2}
}
}
=
\frac{1}{2^{N_{\mathrm{s}}}}
\,
\sum_{ s^{(1)} }\,
\sum_{ s^{(2)} }\,
\ket
{
s^{(1)}; s^{(2)}
}
\,,
\end{split}•
\end{equation}•
which is the tensor product of two trivial $\mathbb{Z}_{2}$ paramagnets.

Eq.~(\ref{eq: Z2 times Z2 paramagnet ground state}) 
corresponds to the choice of ground state in the 
``high temperature" limit $\beta = 0$ and the SPT ground states can be obtained
from the trivial one,
Eq.~(\ref{eq: Trivial Z2 times Z2 paramagnet ground state}),
via the unitary transformation 
$
\mathbb{W}^
{
2 \mathrm{d}
}
_
{
\mathbb{Z}_{2}\times\mathbb{Z}_{2}
}
$, 
whose action on the many-body basis 
$
\{\,
\ket
{
s^{(1)}; s^{(2)}
}
\,\}
$
reads
\begin{widetext}
\begin{equation}
\label{eq: Phase factor in the Z2 times Z2 SPT state in 2 dimensions}
\mathbb{W}^
{
2 \mathrm{d}
}
_
{
\mathbb{Z}_{2}\times\mathbb{Z}_{2}
}
\,
\ket
{
s^{(1)}; s^{(2)}
}
\equiv
e^
{
i\,
W^
{
2 \mathrm{d}
}
_
{
\mathbb{Z}_{2}\times\mathbb{Z}_{2}
}
(s^{(1)};s^{(2)})
}
\,
\ket
{
s^{(1)}; s^{(2)}
}
=
e^
{
i\,\pi\,
\left[
p_{1}\,L(s^{(1)})
+
p_{2}\,L(s^{(2)})
+
p_{12}\,L(s^{(12)})
\right]
}
\,
\ket
{
s^{(1)}; s^{(2)}
}
\,.
\end{equation}•
\end{widetext}
In Eqs.~(\ref{eq: Z2 times Z2 paramagnet ground state}) 
and~(\ref{eq: Phase factor in the Z2 times Z2 SPT state in 2 dimensions}), in order to simplify the notation, we omit the indices $p_{1}$, $p_{2}$ and $p_{12}$ on which the SPT ground states and the unitary transformation depend upon.

\subsubsection
{
Parent Hamiltonians of the 
$\mathbb{Z}_{2}\times\mathbb{Z}_{2}$
state
in 
$
\mathrm{d}
=
2
$
}

\label
{
sec: 
Parent Hamiltonians of the 
Z2 times Z2 SPT state in d=2
}

The Hamiltonian
\begin{equation}
\label{eq: Trivial Hamiltonian of the Z2 times Z2 SPT state}
\begin{split}
H^{2 \mathrm{d}}_{\mathbb{Z}_{2}\times\mathbb{Z}_{2}}
=
\omega_{0}
\,
\sum_{j}\,
\left[
\left(
1
-
X^{(1)}_{j}
\right)
+
\left(
1
-
X^{(2)}_{j}
\right)
\right]
\end{split}•
\end{equation}•
describes a trivial paramagnetic system and has 
Eq.~(\ref{eq: Trivial Z2 times Z2 paramagnet ground state}) as its ground state.
It is the sum of two RK Hamiltonians of the form 
Eq.~(\ref{eq: Definition of the RK Hamiltonian}) 
with $\beta = 0$,
$
\omega_{s,s'}
=
\omega_{0}
>
0
$,
where the sum in 
Eq.~(\ref{eq: RK Hamiltonian})
extends over pairs of states $(s,s')$ which differ by a single spin flip.

Then applying the unitary transformation
$
\mathbb{W}^
{
2 \mathrm{d}
}
_
{
\mathbb{Z}_{2}\times\mathbb{Z}_{2}
}
$
in
Eq.~(\ref{eq: Phase factor in the Z2 times Z2 SPT state in 2 dimensions})
to the Hamiltonian 
Eq.~(\ref{eq: Trivial Hamiltonian of the Z2 times Z2 SPT state})
yields the microscopic model realizing the $8$ classes of 
$
\mathbb{Z}_{2}\times\mathbb{Z}_{2}
$
SPT paramagnets:

\begin{widetext}
\begin{equation}
\label{eq: Z2 times Z2 SPT Hamiltonian}
\begin{split}
\mathcal{H}^{2 \mathrm{d}}_{\mathbb{Z}_{2} \times \mathbb{Z}_{2}}
&\,=
\mathbb{W}^{2 \mathrm{d}}_{\mathbb{Z}_{2} \times \mathbb{Z}_{2}}
\,
H^{2 \mathrm{d}}_{\mathbb{Z}_{2} \times \mathbb{Z}_{2}}
\,
\left(
\mathbb{W}^{2 \mathrm{d}}_{\mathbb{Z}_{2} \times \mathbb{Z}_{2}}
\right)^{-1}
\\
&\,=
\omega_{0}\,
\sum_{j}\,
\Big\{
1
-
X^{(1)}_{j}\,
e^
{
i\,\pi\,
\left(
p_{1} +p_{12}
\right)
+
i\,\frac{\pi}{4}\,
\sum_{< \ell \, \ell' ; \, j >}\,
\left[
p_{1}\,
\left(
1 - Z^{(1)}_{\ell}\,Z^{(1)}_{\ell'}
\right)
+
p_{12}\,
\left(
1 - Z^{(12)}_{\ell}\,Z^{(12)}_{\ell'}
\right)
\right]
}
\Big\}
\\
&\,
+
\omega_{0}\,
\sum_{j}\,
\Big\{
1
-
\,X^{(2)}_{j}\,
e^
{
i\,\pi\,
\left(
p_{2} +p_{12}
\right)
+
i\,\frac{\pi}{4}\,
\sum_{< \ell \, \ell' ; \, j >}\,
\left[
p_{2}\,
\left(
1 - Z^{(2)}_{\ell}\,Z^{(2)}_{\ell'}
\right)
+
p_{12}\,
\left(
1 - Z^{(12)}_{\ell}\,Z^{(12)}_{\ell'}
\right)
\right]
}
\Big\}
\,,
\end{split}•
\end{equation}•
\end{widetext}
where the notation $< \ell \, \ell' ; \, j >$ denotes that the summation in the exponent runs over pairs of nearest neighbors $\ell$ and $\ell'$, which themselves belong to the $6$ nearest sites around a given site $j$ of the triangular lattice.
When $p_{12} = 0$ and $p_{1}=p_{2} = 1$,  
Eq.~(\ref{eq: Z2 times Z2 SPT Hamiltonian}) describes, up to an overall energy shift, two decoupled Levin-Gu Hamiltonians.~\cite{Levin-2012}
On the other hand, the case $p_{12} = 1$ 
encodes strong entanglement between the two Ising systems.


\subsubsection
{
Edge symmetry transformations of the
$\mathbb{Z}_{2}\times\mathbb{Z}_{2}$
state
in 
$
\mathrm{d}
=
2
$
}

The physical properties of the edge of two dimensional bosonic SPT states can be understood from the chiral action of the symmetry on the edge modes~\cite{Lu-2012,Chen-2012-b,Levin-2012,Santos-2014,J-Wang-2014,Else-2014}
which, on a microscopic scale, originates from a non-onsite symmetry
realization on the edge degrees of freedom.~\cite{Chen-2012-b,Levin-2012,Santos-2014,J-Wang-2014}
For the
$\mathbb{Z}_{2} \times \mathbb{Z}_{2}$ SPT states 
considered here, the non-onsite symmetry transformations on the edge
were shown in Ref.~\onlinecite{J-Wang-2014} to be a manifestation 
of the non-trivial type-II cocycles of group cohomology.

The ground state Eq.~(\ref{eq: Trivial Z2 times Z2 paramagnet ground state}) is a simple product state and, as such, the edge states are disentangled from the bulk.
On the other hand, in the other SPT ground states whose expansion coefficients depend upon the domain wall configurations, the effect of flipping an edge spin can change the number of loops $L$, implying an entanglement between edge and bulk degrees of freedom. Therefore by studying the properties of the edge states, we can gain information about the nature of the bulk states and vice versa.

We have found, working directly with the bulk SPT wave functions
Eq.~(\ref{eq: Z2 times Z2 paramagnet ground state}), that the projection of the symmetry transformations Eq.~(\ref{eq: Z2 times Z2 symmetry}) onto the
boundary spins is
\begin{widetext}
\begin{subequations}
\label{eq: Z2 times Z2 symmetry transformation projected on the edge}
\begin{equation}
\begin{split}
\widehat{S}^{(1)}_{\mathbb{Z}_{2},\textrm{edge}}
=
&\,
\prod_{j}\,
X^{(1)}_{j}\,
e^
{
i\,\pi\,
\left(
p_{1} + p_{12}
\right)
+
i\,\frac{\pi}{4}\,
\left[
p_{1}
\left(
1
-
Z^{(1)}_{j}\,Z^{(1)}_{j+1}
\right)
+
p_{12}
\left(
1
-
Z^{(12)}_{j}\,Z^{(12)}_{j+1}
\right)
\right]
}
\,,
\end{split}•
\end{equation}•
\begin{equation}
\begin{split}
\widehat{S}^{(2)}_{\mathbb{Z}_{2},\textrm{edge}}
=
&\,
\prod_{j}\,
X^{(2)}_{j}\,
e^
{
i\,\pi\,
\left(
p_{2} + p_{12}
\right)
+
i\,\frac{\pi}{4}\,
\left[
p_{2}
\left(
1
-
Z^{(2)}_{j}\,Z^{(2)}_{j+1}
\right)
+
p_{12}
\left(
1
-
Z^{(12)}_{j}\,Z^{(12)}_{j+1}
\right)
\right]
}
\,,
\end{split}•
\end{equation}•
\end{subequations}•
\end{widetext}
where the product in
Eq.~(\ref{eq: Z2 times Z2 symmetry transformation projected on the edge})
is taken over the spins on the edge with periodic boundary conditions.
Remarkably these are the same non-onsite symmetry transformations studied in Ref.~\onlinecite{J-Wang-2014}, which shows that 
Eq.~(\ref{eq: Z2 times Z2 paramagnet ground state}) 
gives the correct representation of the ground state
of the two dimensional bosonic SPT states protected by $\mathbb{Z}_{2} \times \mathbb{Z}_{2}$ symmetry.
As is clear from 
Eq.~(\ref{eq: Z2 times Z2 symmetry transformation projected on the edge}),
the non-onsite form acquired by the edge symmetry in the non-trivial SPT phase
implies that domain wall configurations on the edge carry projective representation
of the $\mathbb{Z}_{2}$ symmetry, i.e., 
upon flipping any given spin configuration twice, 
there is a factor of 
$
(-1)^{p}
$
for every domain wall.
Since the total number of domain walls is even,
it follows that 
Eq.~(\ref{eq: Z2 times Z2 symmetry transformation projected on the edge})
is a faithful representation of the $\mathbb{Z}_{2}$ symmetry on the edge.

\section{Conclusions and future directions}
\label{sec:Conclusions and future directions}

We have demonstrated that Rokhsar-Kivelson models offer a useful framework
for constructing microscopic models whose ground states encode the universal properties of bosonic symmetry-protected topological states. 
Although we have illustrated our construction for one- and two-dimensional
models and for certain discrete symmetries (unitary and anti-unitary), we believe our construction should hold true generically for any number of dimensions and for continuous symmetries as well.
We close with a few observations:

\begin{enumerate}
\item
It will certainly be interesting to realize parent Hamiltonians for bosonic SPT states in 
two dimensions with discrete symmetry $\mathbb{Z}_{n}$, for $n \geq 3$. 
In contrast to the $\mathbb{Z}_{2}$ case, the expansion of the 
paramagnetic state in the ordered basis contains more than one type of
domain walls,
suggesting that the coefficients in the expansion of the SPT ground state 
in the $\ket{s}$ basis may depend on orientable loops.
It remains an open question how these domain wall fluctuations can 
coherently give rise to SPT ground states.

\item
We expect that this formalism may offer a path to constructing microscopic models for other classes of bosonic SPT states in three dimensions, as well as interesting examples of bosonic SPT states in two dimensions protected by 
$\mathbb{Z}_{2} \times \mathbb{Z}_{2} \times \mathbb{Z}_{2}$ symmetry, which
are classified by non-trivial Type-III cocycles.~\cite{J-Wang-2014}

\item
A refinement of the group cohomology classification in terms of cobordisms
has been recently proposed to describe three dimensional bosonic SPT states
with time-reversal symmetry.~\cite{Kapustin-2014}
It could be worthwhile to study these time-reversal symmetric phases of matter
using the formalism of RK Hamiltonians, in particular, to understand 
if there is any special meaning attached to the global unitary transformation 
that connects the trivial and the non-trivial SPT ground states,
which could shed new light on the reason behind the 
inability of the group cohomology classification
to account for these phases of matter.
\end{enumerate}

\section*{Acknowledgments}
We thank Eduardo Fradkin for constructive comments on the manuscript
and we acknowledge a useful discussion with Zheng-Cheng Gu
about Ref.~\onlinecite{Levin-2012}.
Research at the Perimeter Institute is supported by the Government
of Canada through Industry Canada and by the Province of Ontario
through the Ministry of Economic Development and Innovation.

\appendix

\section{$\mathbb{Z}_{n}$ operators}
\label{sec: Zn operators}

At each site $j$ of a lattice we consider operators
$
\sigma_{j} 
$
and
$
\tau_{j}
$
represented by $n \times n$ matrices
\begin{equation}
\sigma_{j} 
=
\begin{pmatrix} 
1 & 0 & 0 & 0 \\
0 & \omega & 0 & 0\\  
0 & 0 & \ddots  & 0\\  
0 & 0 & 0 & \omega^{n-1}
\end{pmatrix}
\,,
\end{equation}•
\begin{equation}
\tau_{j} 
=
\begin{pmatrix} 
0 & 0 & 0 & \dots &0& 1  \\
1 & 0 & 0 & \dots &0& 0 \\
0 & 1 & 0 & \dots &0& 0 \\
0 & 0 & 1 & \dots &0& 0 \\
\vdots &0 & 0 & \dots &1 & 0 
\end{pmatrix} 
\,,
\end{equation}•
satisfying
\begin{equation}
\label{eq: Zn algebra appendix}
\begin{split}
\s^{n}_{j}
=
\tau^{n}_{j}
=
1
\,,
\quad
\tau^{\dagger}_{j}\,
\s_{j}\,
\tau_{j}
=
\omega\,
\s_{j}
\,,
\end{split}•
\end{equation}•
where
$
\omega
=
e^
{
i\,\frac{2\,\pi}{n}
}
$
and
$
\bar{\omega}
=
e^
{
-i\,\frac{2\,\pi}{n}
}
$.

The following identities hold for $p = 0,...,n$:
\begin{equation}
\label{eq: Algebra Zn operators}
\begin{split}
\exp
{
\left\{
i\,\frac{2\,\pi p}{n}\,
\left[
\frac{n-1}{2}
+
\sum^{n-1}_{a=1}\,
\frac
{
\left(
\s^{\dagger}_{j}\,\s_{j'}
\right)^{a}
}
{
\bar{\omega}^{a} - 1
}
\right]
\right\}
}
=
\left(
\s^{\dagger}_{j}\,\s_{j'}
\right)^{p}
\,,
\end{split}•
\end{equation}•

\begin{equation}
\label{eq: Identity Zn operators}
\begin{split}
\exp
{
\left\{
-i\,\frac{2\,\pi p}{n}\,
\left[
\frac{n-1}{2}
+
\sum^{n-1}_{a=1}\,
\frac
{
\left(
\s^{\dagger}_{j}\,\s_{j'}
\right)^{a}
}
{
\omega^{a} - 1
}
\right]
\right\}
}
=
\left(
\s^{\dagger}_{j}\,\s_{j'}
\right)^{p}
\,.
\end{split}•
\end{equation}•

\end{document}